\documentclass[fleqn,usenatbib]{mnras}

\usepackage{newtxtext,newtxmath}
\usepackage[T1]{fontenc}

\DeclareRobustCommand{\VAN}[3]{#2}
\let\VANthebibliography\thebibliography
\def\thebibliography{\DeclareRobustCommand{\VAN}[3]{##3}\VANthebibliography}

\usepackage{graphicx}	% Including figure files
\usepackage{amsmath}	% Advanced maths commands
\title[Hyper-runaway and hypervelocity WDs in \textit{Gaia} DR3]{Hyper-runaway and hypervelocity white dwarf candidates in \textit{Gaia} Data Release 3: possible remnants from Ia/Iax supernova explosions or dynamical encounters}

\author[Igoshev A.P. et al.]{
Andrei P. Igoshev,$^{1}$\thanks{E-mails: \href{matilto:a.igoshev@leeds.ac.uk}{a.igoshev@leeds.ac.uk},  \href{mailto:ignotur@gmail.com}{ignotur@gmail.com}}
Hagai Perets,$^{2}$
Na'ama Hallakoun$^{3}$
\\
$^{1}$Department of Applied Mathematics, University of Leeds, LS2 9JT Leeds, UK\\
$^{2}$Physics Department, Technion - Israel Institute of Technology, Haifa 3200002, Israel\\
$^{3}$Department of particle physics and astrophysics, Weizmann Institute of Science, Rehovot 7610001, Israel
}

\date{Accepted XXX. Received YYY; in original form ZZZ}

\pubyear{2022}

\begin{document}
\label{firstpage}
\pagerange{\pageref{firstpage}--\pageref{lastpage}}
\maketitle

\begin{abstract}
Type Ia and other peculiar supernovae (SNe) are thought to originate from the thermonuclear explosions of white dwarfs (WDs). Some of the proposed channels involve the ejection of a partly exploded WD (e.g. Iax SN remnant) or the companion of an exploding WD at extremely high velocities ($>400$ km\,s$^{-1}$). Characterisation of such hyper-runaway/hypervelocity (HVS) WDs might therefore shed light on the physics and origins of SNe. Here we analyse the \textit{Gaia} DR3 data to search for HVS WDs candidates, and peculiar sub-main-sequence (sub-MS) objects. 
We retrieve the previously identified HVSs, and find 46 new HVS candidates. Among these we identify two new unbound WDs and two new unbound sub-MS candidates. The remaining stars are hyper-runaway WDs and hyper-runaway sub-MS stars.
The numbers and properties of the HVS WD and sub-MS candidates suggest that extreme velocity ejections ($>1000$ km\,s$^{-1}$) can accompany at most a small fraction of type Ia SNe, disfavouring a significant contribution of the D6-scenario to the origin of Ia SNe. The rate of HVS ejections following the hybrid WD reverse-detonation channel could be consistent with the identified HVSs. The numbers of lower-velocity HVS WDs could be consistent with type Iax SNe origin and/or contribution from dynamical encounters. We also searched for HVS WDs related to known SN remnants, but identified only one such candidate. 
\end{abstract}


%%%%%%%%%%%%%%%%%%%%%%%%%%%%%%%%%%%%%%%%%%%%%%%%%%

%%%%%%%%%%%%%%%%% BODY OF PAPER %%%%%%%%%%%%%%%%%%

\section{Introduction}
Most stars in the Galaxy reside in the Galactic disc, and have low peculiar velocities. The typical velocity dispersion of stars in the disc is of a few tens of km\,s$^{-1}$. However, a small fraction of stars and compact objects are known to have far higher velocities of hundreds and even thousands of km\,s$^{-1}$ with respect to local standard of rest (LSR) and could even be unbound to the Galaxy.
Such Galactic hyper-runaway and hypervelocity stars and compact objects have been studied for decades, as the origin of their high peculiar velocities provides potential input on the physical processes which have accelerated these stars. These include stellar evolutionary processes such as natal kicks given to neutron stars \citep{Lyne1994Natur,Arzoumanian2002ApJ,Verbunt2017AA,Igoshev2020MNRAS} and possibly other type of compact objects \citep{Repetto2012MNRAS,Repetto2017MNRAS,ElBadry2018MNRAS}; binary evolution processes, such as supernova kicks in binaries \citep{Brandt1995MNRAS,Igoshev2021MNRAS}, ejecting the companion of the exploding stars at velocities comparable to the orbital velocities at the point of the explosion \citep{Blaauw1961BAN,Hoogerwerf2001AA,Eldridge2011MNRAS,Renzo2019AA}, remnants of Ia/Iax supernova explosion; and stellar dynamical processes in collisional cluster environments, where few-body interactions can give rise to hyper-runaway ejections; or through dynamical interactions with massive black holes that may even give rise to extreme hypervelocity stars ejected at hundreds or thousands of km\,s$^{-1}$ \citep{Hil88}.  

One of the most intriguing aspects of the study of hyper-runaway stars is their potential use in constraining and characterising the still highly debated origins of type I (and in particular type Ia) supernovae originating from thermonuclear explosions of white dwarfs (WDs). Different suggested models for normal and peculiar types of such SNe pointed to various channels for the ejection of hyper-runaway WDs, with distinct predictions. Identifying and characterising the properties, rates and distributions of such WDs can therefore shed light on the origins of thermonuclear SNe.

WDs are formed following the evolution of stars with up to $8-10$~M$_\odot$, and typically above $0.8-0.9$~M$_\odot$. More massive stars explode as core-collapse supernovae and form neutron stars or black holes, while lower-mass stars do not evolve to become WDs during a Hubble time. This is true for single stars, while the evolution of stars in interacting binaries could be somewhat altered by mass-transfer and/or stripping. 

In order to put our work in the context we briefly summarise here different scenarios for SNe Ia/Iax and their rates and expected observational outcomes. At the end of the article we compare these rates with number of hyper-runaway WD candidates found in the \textit{Gaia} database.

\textit{The double degenerate dynamical detonations scenario} was suggested by \cite{Gui+2010,finketal10}. In this scenario two CO WDs with thin surface helium layers remaining from single stellar evolution can lead to a SNe Ia (this scenario is sometimes called the dynamically driven double-degenerate double-detonation, or D6, scenario \citealt{shen2018sub}). After the progenitor stars evolve through single stellar and common-envelope evolution, they eventually form a compact WD-WD binary, whose components are then driven towards increasingly shorter periods by gravitational wave-driven inspiral. The Roche lobe overflow of the donor then provides a helium-rich accretion stream onto the companion. The dynamically formed He-layer and its accretion heat the He surface layer which then experiences a thermonuclear detonation \citep{Gui+2010}. The convergence of the detonation front was suggested to give rise to a second detonation inside the accretor's CO core. This second detonation in the CO core disrupts the accretor and gives rise to a type Ia event\footnote{If the second detonation does not occur, only the first weak explosion occurs, likely leading to a peculiar SN \cite{woosleytaamweaver86,Bil+07}. In \cite{Per+10} and \cite{Zen+22} one of suggested and showed that these could be the progenitors of Ca-rich SNe, the primary is a hybrid HeCO WD. In this case, the companion is disrupted and not ejected as a hyper-runaway WD. Whether the remnant from the primary can be ejected is not clear, and likely require 3D models to explore whether an asymmetric explosion occurs, and give rise to the ejection  of the partially burned primary.}.   The proposed D6 scenario therefore differs from the classic double degenerate scenario in a crucial respect: the donor WD survives the SN Ia event. Specifically, in the D6 scenario, the WD donor is ejected after the detonation of the accretor, at the Keplerian velocity at the point of Roche lobe overflow, a speed of several thousand km\,s$^{-1}$, and generally above 1000 km\,s$^{-1}$.

\textit{The hybrid-WD reverse detonation scenario} was suggested by \cite{Pak+21}. In this scenario a highly He-enriched hybrid HeCO-WD begins transferring mass onto a more massive CO-WD companion, similar to the D6 model mentioned above, leading to a He surface detonation of the CO WD. However, the He-surface detonation fails to induce a detonation in the CO WD, and instead, the nuclear burning front propagates back to the donor hybrid HeCO WD, and the shock leads to its core detonation and disruption, which leaves the primary CO WD intact. The leftover primary WD is ejected at a high velocity, comparable with its Keplerian orbital velocity at the time. Like the D6 scenario, the origin of the velocity excitation is the Keplerian motion in the compact binary at the time of the explosion. In this case it is the secondary, less massive, WD which is disrupted, and therefore the typical ejection velocities of the primary are somewhat lower than the D6 case, typically between $1000-1500$~km\,s$^{-1}$. The ejected WD might be somewhat heated from the surface detonation and polluted by the companion ejecta, but likely not very significantly. The expected ejection rates are of the order of 1~per~cent of the Ia SNe rate \citep{Pak+21}.

\textit{The failed-detonation or weak deflagration model for SNe Iax} was suggested by one of us \citep{Jor+12}.
In this scenario, it is suggested that a CO WD accretes mass from a companion at an appropriate rate that allows it to accumulate mass and eventually reach close to Chandrasekhar mass, at which point it detonates producing a SN. In \cite{Jor+12} (see also \citealt{kro+13}) one of us studied the last stages of the evolution of a near-Chandrasekhar CO WD, and suggested that ignition of nuclear burning might not lead to a full detonation, but may only give rise to a weak asymmetric deflagration, which would then partially burn some of the WD, and eject some of its mass, leading to the production of faint peculiar Ia SNe, which we proposed could explain the origin of type Iax SNe. The weak explosion should leave a somewhat lower mass, likely very hot WD (due to burning and ejection of up to a few 0.1~M$_\odot$ of material) but otherwise intact, and polluted with heavier burning product elements. Such an asymmetric explosion would likely also eject the WD at high velocities of hundreds of km\,s$^{-1}$ possibly up to 500~km\,s$^{-1}$ in the most extreme case modelled. We therefore suggested to search for hyper-runaway WDs with peculiar properties, likely being massive, hot and showing significant pollution by intermediate and iron-elements. The inferred rate of type Iax SNe, is of the order of $20-50$~per~cent of the type Ia SNe rate \citep{Fol+13}, if ultra-faint 2008ha-like SNe are included. The rate is likely lower, of the order of $2-10$~per~cent, if only brighter 2002cx-like SNe are considered as part of this class \citep{Li+11}. 

\textit{Single-degenerate double-detonation} scenario was suggested by \cite{woosleytaamweaver86}. In this scenario, it was suggested that a sufficiently massive WD can accrete and accumulate He from a stripped He-rich stellar companion, which becomes an sdB/O star. After a critical mass is deposited on the surface of the WD, a surface helium ignition may occur, which then triggers the explosion of the CO core
of the WD \citep{woosleytaamweaver86}. Like in the D6 and reverse-detonation scenarios discussed above, the disruption of the exploding WD unbinds the companion which is therefore ejected at velocities comparable to its orbital velocity in the progenitor binary. Unlike the previous channels mentioned above, the companion in this case is a sdB/O star whose size and Roche radius are larger than that of a WD and therefore the orbit of the progenitor binary of the system cannot be as close as that of a double-WD system. \cite{Gei+15} suggested that US-708 could be explained by such a scenario and suggested ejection at velocities exceeding 1000~km\,s$^{-1}$, however \cite{Liu+21} showed in a detailed study that in the relevant case, one can at most achieve 600~km\,s$^{-1}$. More generally, although under extreme conditions one might get very high velocities, in most cases the ejection velocities are limited to a lower range of a few hundred km\,s$^{-1}$ \citep{Men+21}. It is therefore possible that this scenario can explain hyper-runaway sdB/O (which would later evolve to become hyper-runaway WDs). \cite{Neu+22} used population synthesis studies and suggested that the theoretical ejection rate of unbound He-rich stars through this mechanism is two orders of magnitude higher than expected given the single identification of US-708, possibly ruling out this scenario.

Beside these scenarios, the hyper-runaway WD could also be ejected dynamically from globular clusters and  Milky Way centre. Some of these WDs could have been stripped from inspiraling galaxies. Some other WD could in fact be members of binaries and received their large speed due to supernova explosion. We consider these alternative routes in more details in the Discussion section.

The discovery of three hypervelocity\footnote{We adopted the following notation throughout the manuscript. We call a star hypervelocity if it is unbound from the Galaxy. The hyper-runaway star has velocity above our minimum velocity threshold.} WDs in the \textit{Gaia} DR2 catalogue provided potential observational evidence of the ex-companions of sub-Chandrasekhar WDs which underwent SNe Ia explosions in a dynamical detonation variant of the double-degenerate scenario \citep{Shen2018ApJ}. 

%While a dynamical detonation scenario for type Ia SNe \citet{Shen2018ApJ,Bauer2021ApJ} suggests that most type Ia SNe should be accompanied by the ejection of hypervelocity WDs at velocities of $1000-2000$~km\,s$^{-1}$, the hybrid WD reverse-detonation scenario proposed by one of us \citep{Pak+21} could also produce such extreme hypervelocity WDs (though likely extending only up to 1500~km\,s$^{-1}$ or so) at far lower rates ($\sim1$~per cent of the type Ia SNe rate). A completely different channel was suggested by one of the authors for the production of HVS WDs with lower velocities of up to $\sim500$~km\,s$^{-1}$ through partial deflagration of near-Chandrasekhar WDs \citep{Jor+12}. Such WDs might be initially very hot and polluted with intermediate and iron elements from fall-back material from the explosion.

O and B stars are known to have a non-negligible fraction of runaway stars, i.e. stars observed to have particularly high peculiar velocities of above $30-40$~km\,s$^{-1}$, significantly larger than their expected initial velocities at birth \citep{Hoogerwerf2001AA,Eldridge2011MNRAS,Renzo2019AA}. WDs which originate from B-stars could therefore posses such high peculiar velocities. The dynamical perturbation by massive perturbers such as giant molecular clouds and stellar clusters, as well as spiral arms in the Galactic disc can excite the stellar velocities over time, and, thereby older stellar populations show higher velocity dispersions. In particular, WDs formed in the disc, whose age can extend up to the age of the Galaxy, could belong to the oldest populations and have tens of km\,s$^{-1}$ peculiar velocities. A small fraction of them might also be part of the Galactic halo, where the velocity dispersion is far greater and can reach hundreds of km\,s$^{-1}$ up to the Galactic escape velocity.

Here we focus on the hyper-runaway/hypervelocity WD regime, at velocities of hundreds of km\,s$^{-1}$ or above, which are unlikely to be produced through stellar evolution or through most of the dynamical processes discussed above, but could be the result of thermonuclear explosions such as the D6, and the reverse hybrid detonation of failed deflagaration/detonation scenarios. 

We therefore focus on WDs with extreme velocities of typically $>500$~km\,s$^{-1}$, much higher than the velocity dispersion of disc stars, and potentially higher than the Galactic escape velocity. However, since in most cases only data for the tangential velocity are known, we identify the fastest 1000 objects, which effectively give us a lower velocity limit of 400~km\,s$^{-1}$ for the 2D velocities throughout this study. This less conservative cut is made in order to avoid missing potential HVS WD candidates, as well as to find potential candidates produced in failed detonation/deflagration. In addition, we also identify other peculiar candidate hyper-runaways and hypervelocity objects which reside above the WD cooling sequence but significantly below the MS. Since WD remnants of SNe might be heated or affected by the explosion, they might not resemble normal WDs, and therefore a complementary search for such ``peculiar'' objects is also presented here. In particular the objects DR61--DR6-3 (identified by \citealt{Shen2018ApJ}) are not located in the expected region of the WDs, and are not considered to be WDs in our initial criteria, as we discuss below. 

%In the following we present our analysis and findings and then discuss them and summarise. 

Our paper is structured as the following: in Section~\ref{s:known} we summarise the information about known HVS candidates, in Section~\ref{s:analysis} we describe our selection procedure and summarise new candidates. In Section~\ref{s:special_objects} we list and discuss observational properties of our main candidates. In Section~\ref{s:snr_wd} we look for potential association between our HVS WDs and supernova remnants. In Section~\ref{sec:discussion}, we list and discuss all potential scenarios to produce HVS WDs and sdBs. 

\section{Previously characterised HVS WD candidates}
\label{s:known}
Here we discuss the detailed properties of the selected objects, found through a literature search and/or other archival observations. Some of these are further discussed later on, in the context of our analysis and the distribution of the candidate samples.

%\subsection{Previously identified HVS candidates}
\begin{itemize}
    \item \textbf{LSPM J1852+6202}, \textbf{Gaia EDR3 5805243926609660032} and \textbf{LP 398-9} were identified as hypervelocity WD candidates by \cite{Shen2018ApJ}. These are their candidate D6-3, D6-1 and D6-2, respectively.  \cite{Sco+18}, taking a higher tangential velocity cutoff, considered only D6-2 to be a hypervelocity star. \cite{Sco+18} suggested that even this candidate is suspicious because of its relatively poor astrometric quality parameters. Moreover, the low radial velocities of D6-2 and D6-3 also cast doubt on the nature of these candidates being bona fide hypervelocity WDs. The radial velocity of D6-1 is $1200\pm 40$~km\,s$^{-1}$ \citep{Shen2018ApJ}.
    
    \item \textbf{LP40-365} also known as GD 492 is a high proper motion WD with peculiar chemical composition \citep{Rad+18,Raddi2018MNRASb} dominated by intermediate-mass elements. It was suggested as remnant of Iax supernova by \cite{Ven+17}. This WD has high tangential velocity of $497.6\pm 1.1$~km\,s$^{-1}$ \citep{Ven+17}. This WD is slowly rotating with spin period of 8.914~h \citep{Hermes2021ApJ}.
    We identified LP~40-365 among our sub-MS stars, but we did not include it into our final selection because it has a nominal two-dimensional velocity of 454~km\,s$^{-1}$ (below sub-MS cut of 550~km\,s$^{-1}$), nevertheless this velocity is consistent with a Iax remnant and omitted only due our focus on even higher velocity objects. \cite{Raddi2019MNRAS4891489R} discovered three more chemically peculiar, runaway stars. These are stars J1603-6613 (also known as Gaia DR2 5822236741381879040), J1825-3757 (also known as Gaia DR2 6727110900983876096) and J0905+2510 (also known as Gaia DR2 688380457507044864). These have Ne dominated atmospheres with presence of O and Mg, low masses ($\approx 0.2$~M$_\odot$) and ejection velocities around 550-600~km\,s$^{-1}$. The star J1825-3757 is found in our extended search. 

    \item \textbf{LP91-84} is a hot subdwarf which is included in a survey of large proper motion stars \citep{Lepine2005AJ}.
    
    \item \textbf{LP93-21} was studied in detail by \cite{Kawka2020MNRAS}. It is suggested to be an ancient WD merger remnant with a mass of 1.1~M$_{\odot}$. It is a warm carbon-dominated atmosphere DQ WD with a peculiar orbit in the Galaxy. Another team suggested that this WD is a type Iax supernova candidate \citep{Ruffini2019MNRAS}. 

    \item\textbf{US-708} and Hyper-MUCHFUSS candidates \citep{Hirsch2005AA,Tillich2011AA} identified one of the first HVSs and the first sdO HVS. 
    Most of the Hyper-MUCHFUSS candidates  identified by \cite{Zie+17} do not not pass our quality thresholds. Specifically, the parallax measurements uncertainties are US-708: 0.067$\pm$0.204 mas; SDSS~J205030.39-061957.8:  0.17$\pm$ 0.148 mas.  SDSS~J121150.27+143716: 0.0454$\pm$0.1124 mas. SDSS~J123137.56+074621.7 is not included, its parallax 0.2278±0.0954. SDSS~J163213.05+205124.0: 0.2629$\pm$0.0961 mas. Finally, SDSS~J164419.45+452326.7 is not included because its two-dimensional velocity as derived from the proper motion is about 310~km\,s$^{-1}$, which is far below our cut for the two-dimensional velocity.
\end{itemize}

\section{Analysis}
\label{s:analysis}

\subsection{Objects selection: WDs with the largest proper motions in \textit{Gaia} DR3}
We search for hyper-runaway stars in \textit{Gaia} data release 3 \citep{Gaia2016AA,Gaia2022arXiv}.
To do so, we first identify the fastest 1000 objects in the \textit{Gaia} database with colours and magnitudes compatible to WDs using similar magnitude and data quality cuts as \cite{GentileFusillo2021MNRAS}. We should note, however, that although most of the identified objects are consistent with being WDs, some were identified in various other studies to be hot subdwarfs stars of type B/O; these are listed in Table~\ref{tab:candidates} as sdB/sdO, along with the relevant reference.

In order to perform our search we calculate the nominal two-dimensional velocity taking:
\begin{equation}
v' \; [\mathrm{km / s}] = \frac{4.74\;  \mu' \; [\mathrm{mas / year}]}{\varpi' \; [\mathrm{mas}]}    
\end{equation}
where $\mu'$ is the measured proper motion and $\varpi'$ is the measured parallax. In order to exclude objects with large uncertainties, we included in our selection only objects which satisfy the following conditions: (1) $\varpi' / \sigma_\varpi > 4$, i.e. having a relative error of parallax measurement below 0.25; and (2) $\varpi' > 0.25$~mas, i.e. objects with nominal distances that are smaller than 4~kpc. We check the quality of the astrometric solution following the criteria of \cite{Fabricius2021AA}, selecting only objects with (1) renormalised unit weight error (RUWE) < 1.4, (2) \texttt{IPD\_FRAC\_MULTI\_PEAK} < 2, (3) \texttt{IPD\_GOF\_HARMONIC\_AMPLITUDE} < 0.1 and (4) \texttt{ASTROMETRIC\_SIGMA5D\_MAX} < 1.5. We then apply the colour-magnitude cut suggested by \cite{GentileFusillo2021MNRAS}:
\begin{equation}
G_\mathrm{abs} > 6 + 5 (G_\mathrm{BP} - G_\mathrm{RP}).
\end{equation}
Our complete ADQL request can be found in Appendix~\ref{s:adql}. We summarise the essential information about candidates with $v' > 400$~km\,s$^{-1}$ in Table~\ref{tab:candidates}. We search for the names and properties of these objects in the Simbad database \citep{Wenger2000AAS}.

\begin{table*}
\centering
\begin{tabular}{lcrrcccccc}
\hline
Name & Gaia DR3 name & $\varpi' \pm \sigma_\varpi$ & $\mu \pm \sigma_\mu$ & Type & $v_r$ & $v_t$ & $v_\mathrm{corr}$ & Cred. interv. \\
     &               & (mas)                       & (mas~year$^{-1}$)    &      & (km~s$^{-1}$) & (km~s$^{-1}$) & (km~s$^{-1}$)  & (km~s$^{-1}$) \\
\hline
HVUn 1   & 5703888058542880896 & 1.362$\pm$ 0.318 & 207.876 $\pm$ 0.424 & -- & -- & 723.6 & 728.5 & (574, 1770)\\
SDSS J125834.93-005946.1 & 3688712561723372672 & 1.419$\pm$ 0.196 & 211.642 $\pm$ 0.334 & DA(1) & 140.62 & 706.8 & 693.5 & (578, 1030)\\
HVsdBC 1 & 6368583523760274176 & 0.308$\pm$ 0.065 & 36.638 $\pm$ 0.098 & -- & -- & 564.1 & 482.0 & (402, 822)\\
HVWDC 1 & 6416314659255288704 & 1.894$\pm$ 0.385 & 221.716 $\pm$ 0.492 & -- & -- & 555.0 & 527.4 & (448, 1308)\\
HVWDC 2 & 6841322701358236416 & 2.787$\pm$ 0.432 & 321.477 $\pm$ 0.475 & -- & -- & 546.7 & 531.9 & (452, 938)\\
HVWDC 3 & 5808675437975384320 & 2.056$\pm$ 0.368 & 232.483 $\pm$ 0.415 & -- & -- & 536.0 & 508.6 & (442, 1156)\\
LSPM J1731+0331 & 4376935406816933120 & 2.056$\pm$ 0.106 & 231.209 $\pm$ 0.126 & -- & -- & 533.0 & 525.3 & (488, 600)\\
HVWDC 4 & 4739233769591077376 & 1.474$\pm$ 0.214 & 165.588 $\pm$ 0.344 & -- & -- & 532.6 & 524.8 & (434, 814)\\
HVUn 2 & 6640949596389193856 & 1.741$\pm$ 0.158 & 193.175 $\pm$ 0.156 & -- & -- & 525.8 & 501.2 & (458, 668)\\
HVWDC 5 & 4753345692095875328 & 6.489$\pm$ 0.265 & 714.484 $\pm$ 0.354 & -- & -- & 521.9 & 518.0 & (486, 570)\\
HVWDC 6 & 6777159394645734400 & 1.641$\pm$ 0.322 & 169.82 $\pm$ 0.428 & -- & -- & 490.6 & 474.2 & (392, 1040)\\
HVWDC 7 & 855361055035055104 & 17.332$\pm$ 0.093 & 1782.657 $\pm$ 0.108 & -- & -- & 487.5 & 471.9 & (482, 492)\\
EC 20559-3552 & 6778670265357654656 & 0.384$\pm$ 0.067 & 37.951 $\pm$ 0.084 & sdB (2) & -- & 469.0 & 445.8 & (354, 672)\\
HVWDC 8 & 4943575978388814976 & 3.85$\pm$ 0.624 & 380.566 $\pm$ 0.78 & -- & -- & 468.5 & 474.3 & (384, 844)\\
HVWDC 9 & 2463291012727113216 & 3.542$\pm$ 0.389 & 348.218 $\pm$ 0.528 & -- & -- & 465.9 & 449.3 & (398, 636)\\
HVWDC 10 & 1241636356209099264 & 3.871$\pm$ 0.333 & 379.02 $\pm$ 0.439 & -- & -- & 464.2 & 460.5 & (406, 580)\\
HVWDC 11 & 5995439960564759296 & 2.842$\pm$ 0.514 & 274.698 $\pm$ 0.688 & -- & -- & 458.1 & 437.1 & (382, 1110)\\
LSPM J2224+1604 & 2737084320170352256 & 1.754$\pm$ 0.28 & 167.723 $\pm$ 0.43 & -- & -- & 453.2 & 447.2 & (372, 788)\\
HVWDC 12 & 2119975000945142272 & 1.551$\pm$ 0.288 & 148.249 $\pm$ 0.591 & -- & -- & 453.1 & 446.4 & (370, 980)\\
SDSS J123728.64+491302.6 & 1544331701176666624 & 1.042$\pm$ 0.168 & 99.5 $\pm$ 0.19 & DA (3) & -36.0 & 452.5 & 444.7 & (358, 692)\\
LSPM J1345+3431 & 1470682632777169664 & 1.561$\pm$ 0.276 & 148.647 $\pm$ 0.276 & -- & -- & 451.4 & 440.6 & (358, 782)\\
SDSS J124743.35-134351.2 & 3528713077053554432 & 0.37$\pm$ 0.09 & 34.637 $\pm$ 0.123 & -- & -- & 443.3 & 399.4 & (298, 634)\\
HVWDC 13 & 729192473703851264 & 2.059$\pm$ 0.432 & 191.878 $\pm$ 0.577 & -- & -- & 441.8 & 421.1 & (348, 970)\\
FAUST 4434 & 6438915331219654400 & 0.868$\pm$ 0.036 & 80.888 $\pm$ 0.041 & sdOBHe (4) & -- & 441.7 & 415.6 & (410, 482)\\
HVWDC 14 & 3537042874067950336 & 1.334$\pm$ 0.101 & 123.765 $\pm$ 0.143 & -- & -- & 439.6 & 411.9 & (388, 530)\\
HVWDC 15 & 1415765359864865408 & 5.146$\pm$ 0.406 & 475.629 $\pm$ 0.751 & -- & -- & 438.1 & 436.3 & (388, 536)\\
PG 1303+122 & 3737057611255721472 & 0.395$\pm$ 0.059 & 35.903 $\pm$ 0.103 & sdB (5) & -81.0 & 430.9 & 394.2 & (326, 544)\\
HVsdBC 2 & 3195038476578336256 & 0.325$\pm$ 0.078 & 29.504 $\pm$ 0.092 & -- & -- & 430.0 & 430.8 & (282, 574)\\
HVWDC 16 & 4615529846653846016 & 3.48$\pm$ 0.383 & 311.918 $\pm$ 0.749 & -- & -- & 424.8 & 401.1 & (364, 586)\\
HVsdBC 3 & 6670029411202563584 & 0.322$\pm$ 0.07 & 28.597 $\pm$ 0.097 & -- & -- & 420.5 & 410.1 & (302, 660)\\
HVWDC 17 & 4925179671389315968 & 1.668$\pm$ 0.318 & 147.564 $\pm$ 0.413 & -- & -- & 419.3 & 402.8 & (332, 804)\\
HVWDC 18 & 2914272062095015552 & 1.98$\pm$ 0.336 & 174.78 $\pm$ 0.444 & -- & -- & 418.3 & 416.6 & (342, 798)\\
LSPM J1240+6710 & 1682129610835350400 & 2.36$\pm$ 0.119 & 208.248 $\pm$ 0.193 & DS (6) & -- & 418.3 & 412.2 & (382, 468)\\
SDSS J123800.09+194631.4 & 3948319763985443200 & 0.452$\pm$ 0.099 & 39.747 $\pm$ 0.14 & D (7) & -69.0 & 416.5 & 397.3 & (286, 568)\\
Ton S 145 & 2335322500798589184 & 0.419$\pm$ 0.082 & 36.657 $\pm$ 0.094 & sdBHe1 (8) & -- & 415.1 & 404.9 & (292, 548)\\
HVUn 3 & 2654214506741818880 & 3.397$\pm$ 0.631 & 297.383 $\pm$ 1.041 & -- & -- & 414.9 & 402.8 & (338, 920)\\
HVWDC 19 & 3611573712136684928 & 2.067$\pm$ 0.44 & 180.325 $\pm$ 0.615 & -- & -- & 413.5 & 391.0 & (330, 1002)\\
HVUn 4 & 4026695083122023552 & 4.122$\pm$ 0.816 & 358.18 $\pm$ 1.118 & -- & -- & 411.8 & 406.3 & (334, 986)\\
HVWDC 20 & 6414789778364569216 & 2.069$\pm$ 0.226 & 178.333 $\pm$ 0.307 & -- & -- & 408.5 & 381.6 & (350, 562)\\
PG 1608+374 & 1378348017099023360 & 0.268$\pm$ 0.049 & 23.111 $\pm$ 0.08 & sdOHe (1) & -- & 408.3 & 383.2 & (290, 514)\\
HVWDC 21 & 2497775064628920832 & 3.719$\pm$ 0.302 & 318.334 $\pm$ 0.426 & -- & -- & 405.7 & 391.3 & (358, 500)\\
HVWDC 22 & 5142197118950177280 & 13.036$\pm$ 0.097 & 1111.31 $\pm$ 0.109 & -- & -- & 404.1 & 400.5 & (398, 410)\\
HVWDC 23 & 1217609832414369536 & 7.336$\pm$ 0.759 & 624.631 $\pm$ 1.191 & -- & -- & 403.6 & 397.2 & (348, 544)\\
2MASS J12564352-6202041 & 5863122429179888000 & 13.237$\pm$ 0.326 & 1124.303 $\pm$ 0.421 & L (9) & -- & 402.6 & 385.2 & (384, 424)\\
HVWDC 24 & 3905910019954089856 & 3.291$\pm$ 0.332 & 279.459 $\pm$ 0.478 & -- & -- & 402.5 & 404.4 & (346, 530)\\
HVWDC 25 & 1212348119518459392 & 11.217$\pm$ 0.353 & 951.562 $\pm$ 0.54 & -- & -- & 402.1 & 399.8 & (380, 430)\\
\hline
\end{tabular}
\caption{The properties of HVS WDs. The units for the velocities are km\,s$^{-1}$; $v_r$ is the radial velocity, $v_t$ is transversal velocity and $v_\mathrm{corr}$ is the transversal velocity corrected for rotation of the Milky Way and for \textit{Gaia} parallax zero offset. In this correction we assume that $R_\odot = 8.34$~kpc, $v_\mathrm{circ} = 240$~km\,s$^{-1}$, and the components of the peculiar solar velocity are $U = 11.1$~km\,s$^{-1}$, $V = 12.24$~km\,s$^{-1}$ and $W = 7.25$~km\,s$^{-1}$, which corresponds to works by \protect\cite{Reid2014ApJ} and \protect\cite{Schoenrich2010MNRAS}. Values in the last column corresponds to $95$~per~cent credible interval for the transversal velocity without correcting for the Milky Way rotation. The priors for velocity and distances are specified in Appendix~\ref{a:posterior}. The stellar types references are: (1) \protect\cite{Kepler2015MNRAS}, (2) \protect\cite{ODonoghue2013MNRAS}, (3) \protect\cite{Eisenstein2006ApJS}, (4) \protect\cite{Geier2017AA}, (5) \protect\cite{Green1986ApJS}, (6) \protect\cite{Kep+16}, (7) \protect\cite{Brown2013ApJ}, (8) \protect\cite{Lamontagne2000AJ} and (9) \protect\cite{Smith2018MNRAS}.}
\label{tab:candidates}
\end{table*}

\begin{table}
    \centering
    \begin{tabular}{llccc}
    \hline
    Identificator & Type of hyper-runaway stars          & Number \\
    \hline
    \multicolumn{2}{l}{Newly identified  WDs and sdBs }  & 32 \\
    HVWDC  & ... of which WDs                            & 25 \\
    HVsdBC & ... of which sdBs                           & 3 \\
    HVUn   & ... of which unknown nature                 & 4 \\
    \hline
    HVsMSC  & Newly identified  sub-MS                   & 14\\ 
            & ... of which unbound (hypervelocity)       & 2 \\  
    \hline
    \end{tabular}
    \caption{Number of newly found hyper-runaway stars.}
    \label{tab:numberCand}
\end{table}

In this search we identified for the first time 32 new hyper-runaway WD and sdB candidates. We give these candidates composite names with an ``HV'' prefix (standing for high-velocity) followed by the object type and their sequential number. We summarise the number of new found candidates in Table~\ref{tab:numberCand}. We explain how we assign the candidate types later on.

It is known that quasars in \textit{Gaia} EDR3 have an average parallax of $-17$~$\mu$as \citep{Lindegren2021AA}. This value varies with the magnitude and the location of the star and is summarised in the \texttt{Python} package \texttt{gaiadr3-zeropoint}\footnote{https://pypi.org/project/gaiadr3-zeropoint/}. It was shown in previous works \citep{Marchetti2019MNRAS,Marchetti2021MNRAS} that accounting for the zero-point offset has a significant impact on the number of stars which appear to be unbound to the Galaxy. The effect is relatively small in our work because we limit the minimal parallax to $0.250$~mas. We can estimate the maximum amplitude of the change as $(17/250) \cdot 450 \approx 30$~km\,s$^{-1}$. We compute velocities corrected for the motion of the local standard of rest and zero point in Table~~\ref{tab:candidates} as $v_\mathrm{corr}$. The maximum change due to the zero-point offset is around 10~km\,s$^{-1}$.

The parallax is measured with significantly worse precision in comparison to the proper motion for the majority of these candidates. Therefore, parallax uncertainty contributes significantly to the uncertainty of the two-dimensional velocity. The parallax measurement is subject to the Lutz-Kelker bias \citep{LutzKelker1973PASP}. Two factors contribute to this bias: (1) a symmetric normal distribution for the parallax errors translates into a right-skewed error distribution for the distances, and (2) there are more stars at larger distances from the Sun in comparison to smaller distances, thus in any parallax-limited sample, the distances are more likely to be underestimated. Since the two-dimensional velocity is proportional to the distance, the nominal velocities in our parallax-limited sample are also expected to be underestimated. Fortunately, there are known mitigation techniques to deal with the Lutz-Kelker bias \citep{BailerJones2015PASP,Igoshev2016AA}. We write the Bayesian posterior for the two-dimensional velocities. The posterior estimate for the velocity is a multiplication of the conditional probabilities to measure a parallax given a distance, the conditional probabilities to measure components of proper motion given a velocity and priors for the velocity and the distance. For the distance we use the Galactic prior suggested by \cite{Verbiest2012ApJ}. For the velocity we use a prior composed of two multiplied normal distributions with $\sigma = 1000$~km\,s$^{-1}$. The details of this calculation are summarised in Appendix~\ref{a:posterior}. We estimate  the 95~per~cent credible interval for each object and provide them in the last column of Table~\ref{tab:candidates}. We make our code calculating posterior velocities publicly available\footnote{https://pypi.org/project/post-velocity/}.

\begin{figure}
	\includegraphics[width=\columnwidth]{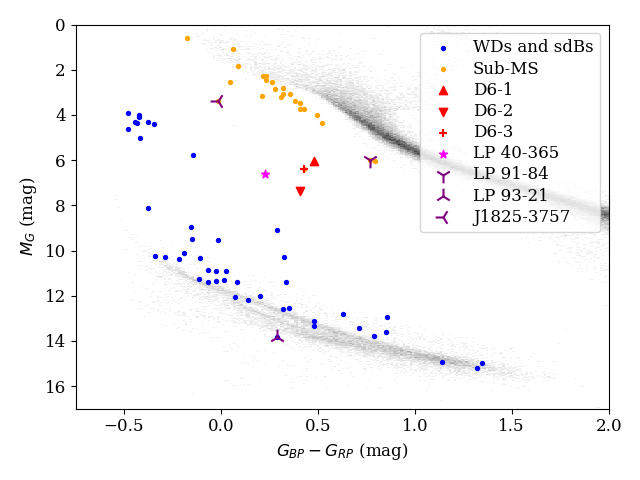}
    \caption{Hertzsprung–Russell diagram for white dwarfs, sdBs and sub-MS candidates with the largest nominal two-dimensional velocities. The \textit{Gaia} 100\,pc sample is plotted in grayscale for reference. }
    \label{fig:hr}
\end{figure}

\subsection{Distribution of HVS candidates in the Hertzsprung-Russell diagram}

We plot the locations of the HVS candidates on the Hertzsprung-Russell diagram in Figure~\ref{fig:hr}. There are three different classes of sources present at this diagram: (1) WD-like sources coincide or are slightly above the WD sequence, (2) hot subdwarf B-like sources are concentrated around $G\approx 4$ and $G_\mathrm{Bp} - G_\mathrm{Rp}\approx -0.4$ and (3) sources located between the WD sequence and the main sequence, including DR6-1, DR6-2 and DR6-3 earlier identified by \cite{Shen2018ApJ}. 

Some hot subdwarf O and B stars are known as hyper-velocity sources \citep[see][for a review and references]{Heber2009ARAA}. For example, the sdO star US~708 (not included in our sample because $\varpi' = 0.0672 < 0.25$~mas) has a radial velocity of $708\pm 15$~km\,s$^{-1}$ \citep{Hirsch2005AA}. Moreover, up to 20~per~cent of sdB stars belong to the Milky Way halo \citep{Napiwotzki2008ASPC}, thus their velocities are expected to be hundreds of km\,s$^{-1}$ with respect to the LSR of the thin disc. We classify candidates with $4 \leq M_G \leq 6$ and $-0.5 \leq G_\textrm{BP}-G_\textrm{RP} \leq -0.25$ as HVsdBC, i.e. high-velocity sdB candidates.

\subsection{Additional data for the HVS WD candidates}
Some additional data exist for other candidates, that were not previously identified as HVS. These objects might be of special interest due to their velocities.

In Table~\ref{tab:mass_and_temperature} we present the estimates of the mass and temperature of our candidate WDs taken from \cite{GentileFusillo2021MNRAS}. In the following we briefly summarise our knowledge about these and other candidates from the \citeauthor{GentileFusillo2021MNRAS} study and other sources. \citet{GentileFusillo2021MNRAS} assigns each WD candidate a parameter, $P_\mathrm{WD}$, indicating its probability of being a WD. We classify stars high $P_\mathrm{WD}$ and fitted WD model atmosphere as HVWDC i.e. hyper-velocity WD candidates. It leaves us with four stars, Gaia DR3 5703888058542880896, 6640949596389193856, 2654214506741818880 and 4026695083122023552, which have a small probability of being WDs ($P_\mathrm{WD} < 0.9$), while their magnitudes and colours seem to be incompatible with sdBs. We classify these as HVUn i.e. high-velocity unknown nature. These objects, that require additional spectral investigation, are among the most interesting sources found in this work.

\begin{itemize}
    \item {LSPM J1240+6710/Gaia~DR3 1682129610835350400}
    \citep{Kep+16} studied this WD due to its unique atmospheric composition, significantly dominated by oxygen. They proposed it is related to an atypical stellar evolution, likely involving a violent very-late thermal pulse during the post-AGB stage. However, such evolution should not provide any velocity kick. \cite{Gan+20} found that this object has high velocity of $\approx 250$~km\,s$^{-1}$ in the direction opposite to the Milky Way rotation.
    
    Instead, here we suggest that our identification of this WD as a HVS WD, together with its unique composition, that includes even traces of Si, could be consistent with the scenario for Iax SN suggested by one of us \citep{Jor+12}. There we proposed that a near-Chandrasekhar WD experiences an explosive asymmetric partial deflagration event which burns only a fraction of the WD, but leaves behind a bound partially-burnt WD remnant, which is ejected at a high velocity (due to the asymmetric explosion). The atmosphere of such a WD would also be polluted by fallback burnt material, potentially consistent with the observations of SDSS~J123800.09+194631.4

    \item SDSS~J015938.43-081242.4 is a WD of type DA \citep{Kilic2006AJ}. 
    %The temperature is estimated using the pure-hydrogen model as $T=7742$~K and mass is $0.47$~M$_{\odot}$. If the atmosphere is assumed to be pure helium, the temperature is $T=7630$~K and mass is 0.47~M$_{\odot}$ according to \citep{Gentile2019MNRAS}.

    \item EC~20559-3552 is classified as a hot subdwarf \citep{ODonoghue2013MNRAS} with a $U-B$ colour of $-1.1$. 
\end{itemize}

\begin{figure}
    \centering
    \includegraphics[width=\columnwidth]{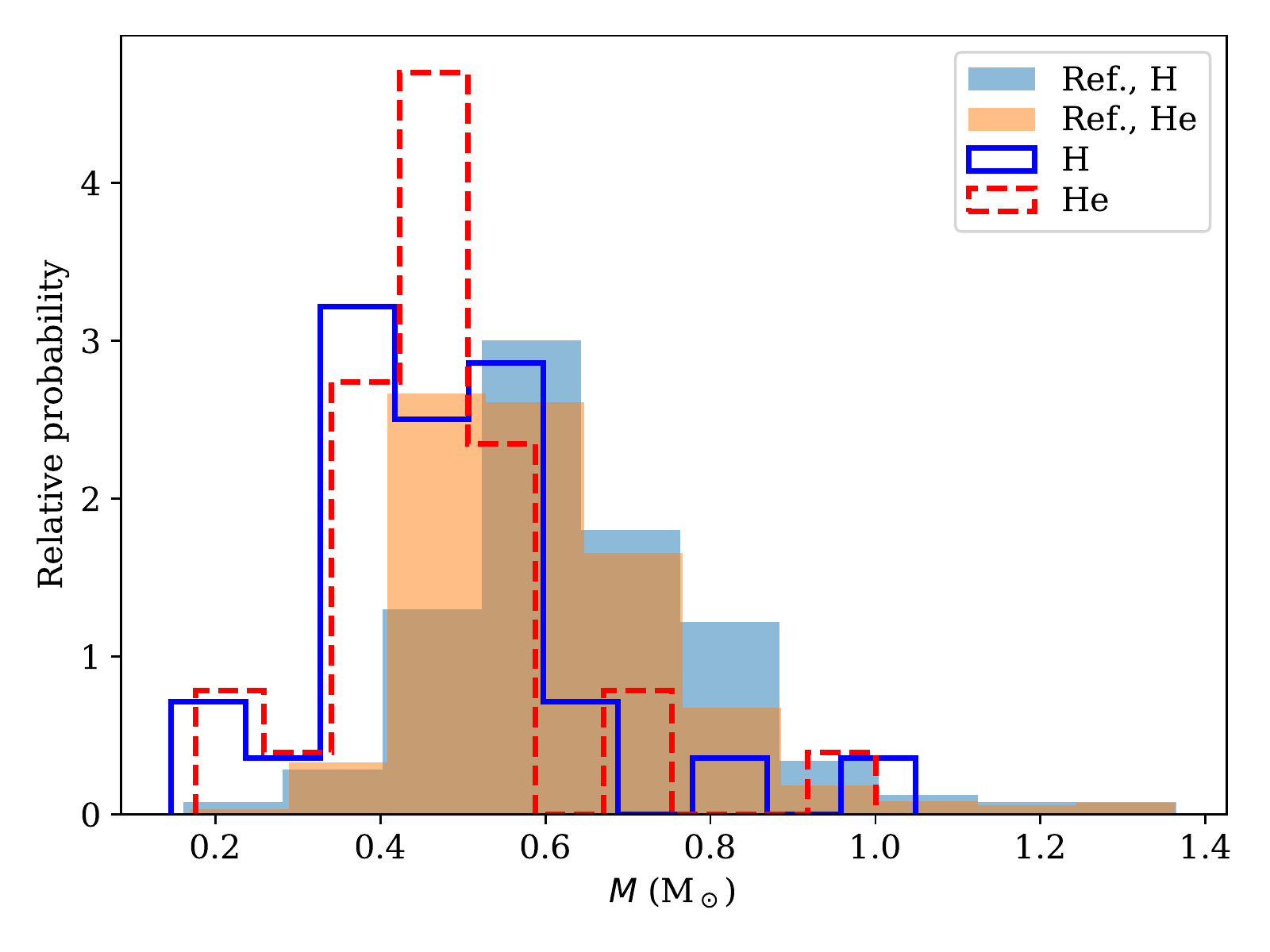}
    \caption{Distribution of derived masses for HVS WD candidates (solid and dashed lines) in comparison to local WD population with $\varpi' > 20$ (filled histograms). Mass measurements are taken from catalogue by \protect\cite{GentileFusillo2021MNRAS}. }
    \label{fig:distr_mass}
\end{figure}

We show mass distribution of candidates with derived mass by \cite{Gentile2019MNRAS} in Figure~\ref{fig:distr_mass}. Our HVS candidates seems to be slightly less massive than local WDs. It might be a result of observational bias, because low-mass WDs have larger radii and thus are expected to be brighter than high-mass WDs. Since our sample is parallax limited, we naturally tend to discover more bright sources.

\begin{table*}
\centering
\begin{tabular}{ccccccccc}
\hline
SDSS & Gaia DR3 & $P_\mathrm{wd}$ & $T_\mathrm{eff,\; H}$ & $M_\mathrm{H}$ & $\chi^2 (\mathrm{H})$ & $T_\mathrm{eff,\; He}$  & $M_\mathrm{He}$ & $\chi^2 (\mathrm{He})$ \\
\hline
 & 5703888058542880896 & 0.447\\
J125834.93-005946.1 & 3688712561723372672 & 0.963 & 11453.8$\pm$1445.1 & 0.146$\pm$0.051 & 0.04 & 12251.4$\pm$1525.0 & 0.176$\pm$0.073 & 0.76\\
 & 6368583523760274176 & 0.011\\
 & 6416314659255288704 & 0.727 & 8880.7$\pm$1232.1 & 0.268$\pm$0.169 & 0.01 & 8866.3$\pm$1362.5 & 0.266$\pm$0.161 & 0.05\\
 & 6841322701358236416 & 0.994 & 12669.3$\pm$2004.7 & 0.462$\pm$0.157 & 2.91 & 12666.3$\pm$1762.8 & 0.463$\pm$0.173 & 5.21\\
 & 5808675437975384320 & 0.996 & 14560.4$\pm$3163.4 & 0.524$\pm$0.242 & 1.41 & 14511.5$\pm$3087.8 & 0.512$\pm$0.276 & 2.85\\
 & 4376935406816933120 & 0.99 & 24160.0$\pm$1473.1 & 0.391$\pm$0.033 & 32.58 & 27730.5$\pm$2018.4 & 0.41$\pm$0.021 & 24.42\\
 & 4739233769591077376 & 0.999 & 25986.7$\pm$6911.6 & 0.603$\pm$0.278 & 0.02 & 30549.4$\pm$10543.3 & 0.68$\pm$0.285 & 0.1\\
 & 6640949596389193856 & 0.237\\
 & 4753345692095875328 & 0.944 & 5626.9$\pm$352.9 & 0.357$\pm$0.098 & 0.79 & 5575.2$\pm$326.2 & 0.354$\pm$0.094 & 0.87\\
 & 6777159394645734400 & 0.998 & 22445.8$\pm$5002.5 & 0.488$\pm$0.208 & 0.61 & 26274.1$\pm$7204.0 & 0.527$\pm$0.203 & 1.16\\
J104559.14+590448.2 & 855361055035055104 & 1.0 & 8720.9$\pm$183.1 & 1.049$\pm$0.025 & 0.29 & 8505.6$\pm$168.4 & 1.0$\pm$0.026 & 0.76\\
 & 6778670265357654656 & 0.008\\
 & 4943575978388814976 & 0.976 & 7076.2$\pm$1410.6 & 0.57$\pm$0.412 & 0.16 & 6952.2$\pm$1403.7 & 0.515$\pm$0.414 & 0.17\\
J015938.43-081242.3 & 2463291012727113216 & 0.966 & 8305.5$\pm$1127.6 & 0.469$\pm$0.247 & 2.1 & 8216.6$\pm$1135.2 & 0.47$\pm$0.227 & 1.96\\
J144205.71+220328.1 & 1241636356209099264 & 0.979 & 8638.9$\pm$938.5 & 0.535$\pm$0.203 & 0.02 & 8464.9$\pm$910.6 & 0.473$\pm$0.174 & 0.05\\
 & 5995439960564759296 & 0.985 & 15469.5$\pm$4435.2 & 0.807$\pm$0.333 & 0.61 & 14078.3$\pm$3745.7 & 0.725$\pm$0.354 & 0.17\\
J222403.94+160405.0 & 2737084320170352256 & 0.997 & 17197.0$\pm$4372.3 & 0.376$\pm$0.148 & 7.91 & 18740.0$\pm$5577.2 & 0.431$\pm$0.2 & 10.11\\
 & 2119975000945142272 & 0.978 & 12084.9$\pm$3110.8 & 0.37$\pm$0.165 & 0.38 & 12363.5$\pm$2868.2 & 0.38$\pm$0.222 & 0.89\\
J123728.64+491302.6 & 1544331701176666624 & 0.969 & 13948.4$\pm$1630.4 & 0.152$\pm$0.061 & 0.87 & 14465.3$\pm$1445.0 & 0.181$\pm$0.068 & 3.41\\
J134503.42+343140.6 & 1470682632777169664 & 0.996 & 14517.6$\pm$2597.4 & 0.349$\pm$0.112 & 0.18 & 14340.6$\pm$2320.8 & 0.361$\pm$0.119 & 0.05\\
J124743.35-134351.2 & 3528713077053554432 & 0.047\\
J103239.96+282724.9 & 729192473703851264 & 0.996 & 13354.5$\pm$3582.8 & 0.527$\pm$0.288 & 1.96 & 13121.5$\pm$3310.7 & 0.486$\pm$0.333 & 3.07\\
 & 6438915331219654400 & 0.012\\
 & 3537042874067950336 & 0.932 & 35775.4$\pm$4798.4 & 0.375$\pm$0.061 & 4.94 & \\
 & 1415765359864865408 & 0.969 & 6112.5$\pm$649.1 & 0.494$\pm$0.219 & 4.4 & 6008.1$\pm$639.5 & 0.48$\pm$0.206 & 4.31\\
J130543.96+115840.8 & 3737057611255721472 & 0.007\\
 & 3195038476578336256 & 0.012\\
 & 4615529846653846016 & 0.853 & 6919.0$\pm$756.9 & 0.351$\pm$0.159 & 0.7 & 6805.0$\pm$740.6 & 0.343$\pm$0.139 & 0.66\\
 & 6670029411202563584 & 0.014\\
 & 4925179671389315968 & 0.991 & 12476.4$\pm$3096.7 & 0.388$\pm$0.186 & 0.13 & 12644.2$\pm$2953.6 & 0.397$\pm$0.237 & 0.0\\
 & 2914272062095015552 & 0.982 & 10992.6$\pm$1495.3 & 0.421$\pm$0.184 & 0.65 & 10976.6$\pm$1641.5 & 0.405$\pm$0.171 & 1.43\\
 & 1682129610835350400 & 0.999 & 22494.9$\pm$4178.4 & 0.514$\pm$0.129 & 0.13 & 25261.6$\pm$7590.0 & 0.536$\pm$0.072 & 0.0\\
J123800.09+194631.4 & 3948319763985443200 & 0.188\\
 & 2335322500798589184 & 0.028\\
J223808.19+003247.6 & 2654214506741818880 & 0.624\\
 & 3611573712136684928 & 0.998 & 14549.2$\pm$4194.3 & 0.514$\pm$0.307 & 1.83 & 14440.2$\pm$4068.7 & 0.498$\pm$0.359 & 3.04\\
J120037.57+320330.7 & 4026695083122023552 & 0.887\\
 & 6414789778364569216 & 0.998 & 17801.3$\pm$3379.0 & 0.464$\pm$0.149 & 0.0 & 17422.9$\pm$4240.0 & 0.464$\pm$0.213 & 0.28\\
J161023.39+371315.9 & 1378348017099023360 & 0.011\\
J024837.53-003123.9 & 2497775064628920832 & 0.973 & 9860.0$\pm$1338.0 & 0.489$\pm$0.211 & 6.77 & 9510.6$\pm$1358.8 & 0.451$\pm$0.168 & 7.86\\
J014809.10-171222.0 & 5142197118950177280 & 0.987 & 7268.0$\pm$130.0 & 0.515$\pm$0.033 & 6.44 & 7138.0$\pm$127.0 & 0.461$\pm$0.011 & 6.11\\
J153719.45+223727.6 & 1217609832414369536 & 0.969 & 4553.5$\pm$568.9 & 0.415$\pm$0.309 & 0.88 & 4626.6$\pm$434.6 & 0.442$\pm$0.272 & 0.89\\
 & 5863122429179888000 & 0.995 & 4561.6$\pm$181.8 & 0.521$\pm$0.093 & 4.88 & 4587.7$\pm$145.0 & 0.512$\pm$0.08 & 4.98\\
J120722.82+091722.3 & 3905910019954089856 & 0.993 & 10812.7$\pm$1731.5 & 0.628$\pm$0.251 & 0.28 & 10846.8$\pm$1947.1 & 0.564$\pm$0.242 & 0.08\\
J151530.71+191130.8 & 1212348119518459392 & 0.958 & 4424.0$\pm$363.3 & 0.401$\pm$0.152 & 0.2 & 4533.9$\pm$271.0 & 0.435$\pm$0.135 & 0.2\\
\hline
\end{tabular}
\caption{Physical properties of hyper-runaway WDs found by \protect\cite{GentileFusillo2021MNRAS}. Here $P_\mathrm{wd}$ is the probability of object to be a WD, 
$T_\mathrm{eff,\; H}$ H and $M_\mathrm{H}$ correspond to temperature and mass estimated if the compositions is pure hydrogen. $T_\mathrm{eff\; He}$  and $M_\mathrm{He}$ correspond to pure helium composition. We also provide two $\chi^2$ values by \protect\cite{GentileFusillo2021MNRAS} for hydrogen and helium atmosphere compositions as $\chi^2 (\mathrm{H})$, $\chi^2 (\mathrm{He})$ respectively. }
\label{tab:mass_and_temperature}
\end{table*}

\subsection{Sub-main-sequence peculiar candidates}
Our strict filtering procedure described in the previous section did not identify some known hyper-runaway objects suspected to be related to WDs ejected following a thermonuclear SN explosion (e.g. the D6-1, D6-2 and D6-3 objects found by \citealt{Shen2018ApJ}). In order to identify these candidates and allow for the possible identification of peculiar objects which might not resemble normal WDs, we select an additional sample of hyper-runaway candidates for which we relax our magnitude selection criteria replacing it with the following:
\begin{equation}
G_\mathrm{abs} > 6.67  (B_p - R_p) + 0.66        
\end{equation}
In this case we select for objects which are positioned just below the main sequence (MS). Our ADQL request for these candidates can also be found in Appendix~\ref{s:adql}. This gives rise to the identification of not only sub-MS objects, but also potential main sequence hyper-runaway stars. Such MS stars are selected although they appear to reside below the MS because their measured parallax is overestimated in comparison to real parallax, thus their absolute magnitude is underestimated. While also of interest, the latter MS candidates are not the focus of our current paper. In order to limit the number of these objects we additionally impose a cut of $v' > 550$~km\,s$^{-1}$ to this sample.
Our final sub-MS candidates are shown in Table~\ref{tab:candidates2}. Newly discovered sources were given a designation HVsMSC, i.e. hyper-runaway sub-MS candidates.

\begin{table*}
\centering
\begin{tabular}{lcrrcccccc}
\hline
Name & Gaia DR3 name & $\varpi' \pm \sigma_\varpi$ & $\mu \pm \sigma_\mu$ & Type & $v_r$ & $v_t$ & $v_\mathrm{corr}$ & Cred. interv. \\
     &               & (mas)                       & (mas~year$^{-1}$)    &      & (km~s$^{-1}$) & (km~s$^{-1}$) & (km~s$^{-1}$)  & (km~s$^{-1}$) \\
\hline
LSPM J1852+6202 / D6-3 & 2156908318076164224 & 0.423$\pm$ 0.099 & 211.996 $\pm$ 0.202 & -- & 20 & 2374.3 & 2313.6 & (1552, 2470)\\
D6-1 & 5805243926609660032 & 0.531$\pm$ 0.07 & 211.749 $\pm$ 0.088 & -- & 1200 & 1890.5 & 1736.8 & (1512, 2370)\\
LP 398-9 / D6-2 & 1798008584396457088 & 1.194$\pm$ 0.065 & 259.514 $\pm$ 0.089 & -- & 20 & 1030.6 & 1024.4 & (938, 1166)\\
SDSS J145847.01+070754.4 & 1160986392332702720 & 0.407$\pm$ 0.032 & 62.771 $\pm$ 0.044 & -- & -117.0 & 730.6 & 656.8 & (630, 852)\\
BPS BS 16470-0087 & 3946876384391994496 & 0.4$\pm$ 0.023 & 59.154 $\pm$ 0.031 & A1.7 (1) & 76.0 & 700.4 & 620.9 & (624, 778)\\
HVsMSC 1 & 3734729567182624512 & 0.333$\pm$ 0.045 & 48.26 $\pm$ 0.075 & -- & -- & 687.0 & 606.1 & (526, 828)\\
J1825-3757 & 6727110900983876096 & 1.051$\pm$ 0.028 & 147.814 $\pm$ 0.033 & -- & -47.0 & 666.5 & 627.0 & (634, 704)\\
HVsMSC 2 & 2316981409896303232 & 0.564$\pm$ 0.035 & 75.776 $\pm$ 0.047 & -- & -- & 636.8 & 584.1 & (566, 720)\\
LP   91-84 & 1063044954547608064 & 1.665$\pm$ 0.022 & 219.672 $\pm$ 0.026 & sdB (2) & -- & 625.6 & 601.1 & (610, 642)\\
HVsMSC 3 & 5814962273679342208 & 0.402$\pm$ 0.023 & 52.708 $\pm$ 0.024 & -- & -- & 621.8 & 525.8 & (564, 708)\\
HVsMSC 4 & 3792840680855962240 & 0.46$\pm$ 0.035 & 60.13 $\pm$ 0.051 & -- & -- & 619.9 & 541.6 & (538, 718)\\
PHL  5459 & 6610315175214347264 & 0.407$\pm$ 0.023 & 52.446 $\pm$ 0.029 & -- & -- & 610.4 & 537.5 & (548, 678)\\
HVsMSC 5 & 1244274913532995072 & 0.26$\pm$ 0.061 & 33.307 $\pm$ 0.068 & -- & -- & 608.0 & 533.6 & (380, 700)\\
HVsMSC 6 & 5114953763436002816 & 0.369$\pm$ 0.027 & 47.071 $\pm$ 0.039 & -- & -- & 604.6 & 529.4 & (524, 694)\\
HVsMSC 7 & 6683685723576804992 & 0.29$\pm$ 0.022 & 36.799 $\pm$ 0.025 & -- & 261.0 & 600.7 & 517.0 & (524, 702)\\
HVsMSC 8 & 4426088459959127552 & 0.274$\pm$ 0.028 & 34.592 $\pm$ 0.037 & -- & -- & 599.2 & 510.4 & (494, 718)\\
HVsMSC 9 & 1480060406106959232 & 0.307$\pm$ 0.019 & 38.322 $\pm$ 0.025 & -- & -- & 592.5 & 528.3 & (520, 660)\\
SDSS J145110.86+335624.2 & 1292695756353196800 & 0.26$\pm$ 0.036 & 32.365 $\pm$ 0.045 & -- & -- & 590.6 & 534.7 & (442, 692)\\
HVsMSC 10 & 2328667912829146368 & 0.385$\pm$ 0.047 & 47.513 $\pm$ 0.056 & -- & -- & 585.7 & 511.7 & (464, 718)\\
EC 00179-6503 & 4900121354714539136 & 0.453$\pm$ 0.016 & 55.444 $\pm$ 0.024 & sdO (3) & 71.0 & 579.8 & 504.9 & (542, 622)\\
HVsMSC 11 & 6849636078710196608 & 0.335$\pm$ 0.024 & 40.165 $\pm$ 0.029 & -- & -- & 568.6 & 484.2 & (500, 660)\\
SDSS J171531.67+271545.5 & 4574342862635194368 & 0.403$\pm$ 0.025 & 48.077 $\pm$ 0.035 & -- & -245.5 & 565.9 & 529.2 & (506, 646)\\
GD 159 & 1244919124267663488 & 0.94$\pm$ 0.019 & 111.472 $\pm$ 0.023 & A (4) & -- & 561.8 & 528.0 & (540, 584)\\
HVsMSC 12 & 3924826739553387648 & 0.468$\pm$ 0.051 & 55.205 $\pm$ 0.065 & -- & -- & 559.5 & 497.7 & (456, 684)\\
HVsMSC 13 & 4937802236674281088 & 0.455$\pm$ 0.072 & 53.262 $\pm$ 0.112 & -- & -- & 554.5 & 519.0 & (418, 734)\\
HVsMSC 14 & 4992603339310680192 & 0.281$\pm$ 0.042 & 32.739 $\pm$ 0.06 & -- & -- & 553.0 & 479.0 & (406, 652)\\
%GD 492 & 1711956376295435520 & 1.638$\pm$ 0.026 & 156.689 $\pm$ 0.041 & b'DZ' & 497.6 & 453.6 & 445.9 & (440, 468)\\
\hline
\end{tabular}
\caption{The properties of hyper-runaway objects. The units for velocities are km\,s$^{-1}$; $v_r$ is the radial velocity, $v_t$ is transversal velocity and $v_\mathrm{corr}$ is the transversal velocity corrected for rotation of the Milky Way and for \textit{Gaia} parallax zero offset. In this correction we assume that $R_\odot = 8.34$~kpc, $v_\mathrm{circ} = 240$~km\,s$^{-1}$, and the components of the peculiar solar velocity are $U = 11.1$~km\,s$^{-1}$, $V = 12.24$~km\,s$^{-1}$ and $W = 7.25$~km\,s$^{-1}$ which correspond to works by \protect\cite{Reid2014ApJ} and \protect\cite{Schoenrich2010MNRAS}. Values in the last column correspond to $95$~per~cent credible interval for transversal velocity without correcting for the Milky Way rotation. Priors for the velocity and distances are specified in Appendix~\ref{a:posterior}. The stellar types references are: (1) \protect\cite{Brown2008AJ}, (2) \protect\cite{Sayres2012AJ}, (3) \protect\cite{Lynn2004MNRAS} and (4) \protect\cite{Greenstein1969ApJ}. The radial velocities of D6-1, D6-2 and D6-3 candidates are from \protect\cite{Shen2018ApJ}.
%and \protect\cite{Ven+17}.}
}
\label{tab:candidates2}
\end{table*}

\section{Objects of special interest and prime targets for follow-up characterisation}
\label{s:special_objects}
Here we first discuss our new prime candidates for follow-up observations, and then briefly discuss candidates in our WD and sub-MS samples, and any studies already done on any of these candidates.

\subsection{Hypervelocity WD candidates}
We find four new sources which have velocities exceeding 700~km\,s$^{-1}$ and are likely (with the caveat of large measurement uncertainties) unbound from the Galaxy, and therefore require a significant velocity kick. Two of these are found in our WD sample, HVUn 1 (Gaia DR3 5703888058542880896) and SDSS~J125834.93-005946.1 / Gaia DR3 3688712561723372672, where SDSS~J125834.93-005946.1 also has a known radial-velocity measurement of 140~km\,s$^{-1}$.
Another two likely unbound hypervelocity sources were found in our second sample of sub-MS HVS candidates: SDSS J145847.01+070754.4 and BPS BS 16470-0087. %Further studies/data indicate they are WDs, and we mention them here, and discuss them in more depth in the next section.

Given their kinematics, all four sources are prime targets for follow-up observations to better characterise their properties.

\subsection{Bound/marginally-bound hyper-runaway WDs}
\cite{Mon+18} find the local escape velocity from the Galaxy to be 580$\pm63$~km\,s$^{-1}$, and suggest it decreases monotonically between 640~km\,s$^{-1}$ at 4~kpc to 550~km\,s$^{-1}$ at 11~kpc (Galactocentric distances). It is therefore likely that our two fastest new HVS WD candidates are unbound hypervelocity WDs kicked following an explosive event, or strong dynamical interaction. The next 8 candidates in Table~\ref{tab:candidates} have tangential velocities ranging between $520-565$~km\,s$^{-1}$. A non-negligible radial-velocity component could potentially make these WDs be unbound hypervelocity WDs, but overall these might be bound WDs, in which case they are likely to be on highly eccentric orbits. In principle, these just might be the extreme tail of normal halo WDs. Distinguishing between these possibilities require knowledge on the radial-velocity component and/or a good age estimate, given that halo WDs are expected to be old. 
We have searched for archival data of radial velocities for these objects but found no additional data. 

Since halo-formed WDs originate from very old ($>10$ Gyr; \citealt{Jof+11,Kil+19}) populations,
identifying younger WDs among these would suggest a disc origin, and hence a large kick, in order to explain their measured velocities. WDs involved in a SN explosion might also have been heated through accretion of material (dynamical detonation in the double-degenerate case; \citealt{Shen2018ApJ}) or a weak deflagration (for the Iax SNe, as we originally suggested; \citealt{Jor+12}), and appear peculiar and/or younger. D6-1--D6-3 for example, have peculiar positions on the HR diagram, suggested to be related to material accretion from the exploding companion \citep{Shen2018ApJ}, while LP~93-21 shows a peculiar composition, suggested to be related to a Iax SN \citep{Ruffini2019MNRAS}, and similarly \citep{Ven+17,Rad+18} for the LP~40-365 object. 

\begin{figure}
	\includegraphics[width=\columnwidth]{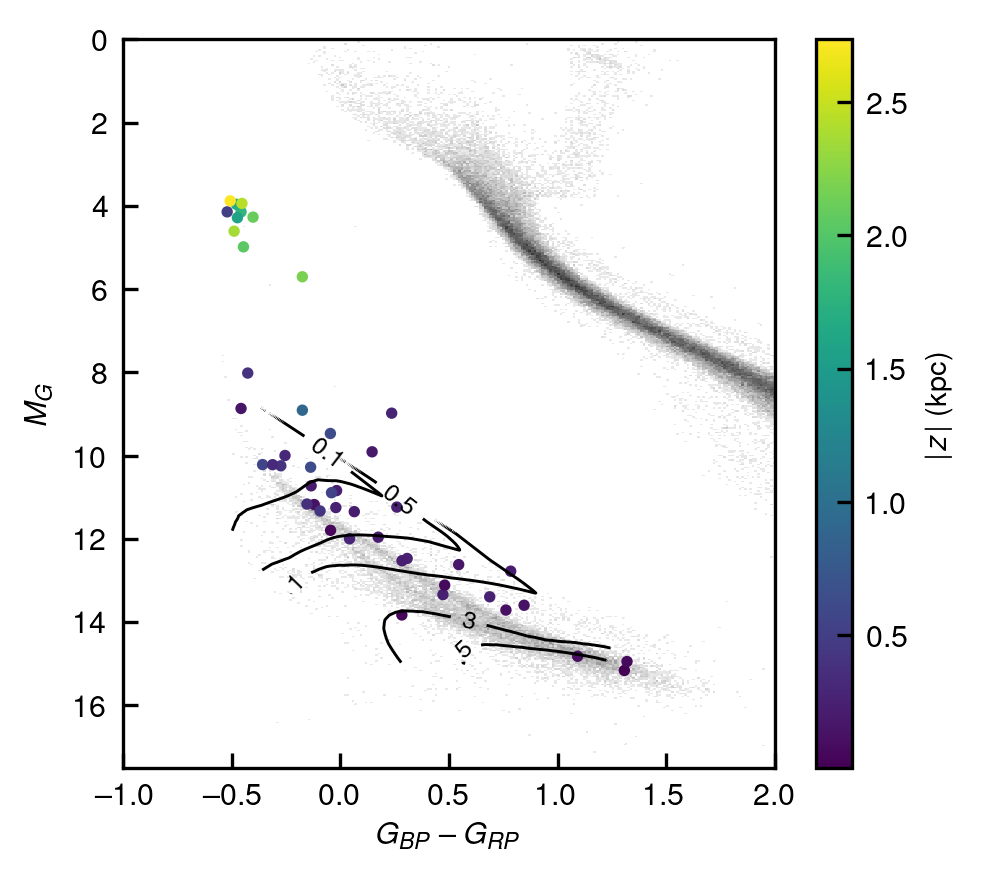}
    \caption{The HVS WD candidate de-reddened \citep[using][]{Capitanio_2017} location on the \textit{Gaia} HR diagram, coloured by the absolute height above/below the Galactic disc. Theoretical cooling-age isochrones for DA WDs \citep{Bedard_2020} are shown in black (the labels mark the cooling age in Gyr). The \textit{Gaia} 100\,pc sample is plotted in grayscale for reference.}
    \label{fig:HRD_CoolingAges}
\end{figure}

For a sub-sample of our candidates we can estimate the WD cooling ages, as well as the total ages (since zero-age MS), using the \texttt{WD\_models} Python package\footnote{\href{https://github.com/SihaoCheng/WD_models}{https://github.com/SihaoCheng/WD\_models}}. However, this approach is limited to more massive WDs. WDs of masses lower than 0.5~M$_\odot$ could not form through normal stellar evolution of single stellar progenitor over a Hubble time. These He or hybrid HeCO WDs \citep{zen+19} have likely undergone a binary evolution stripping process. It is therefore difficult to estimate their true age, in that case, and we cannot exclude a halo origin. To some extent, this could also be the case for slightly more massive WDs which might have been affected by binary evolution, even if their mass is consistent with the age of the Galaxy, in which case their total ages might appear older than they are. The WDs with estimated masses and ages are shown in Table~\ref{tab:mass_and_age}. We plot constant age contours for WDs in Figure~\ref{fig:HRD_CoolingAges}. We also compute the kinematic ages as $b / \mu_b$ for WDs where the sign of proper motion in latitudinal direction coincides with the sign of Galactic latitude. Typical oscillations in the Galactic gravitational potential occur on timescales comparable to 100~Myr, thus even WDs with cooling ages of $0.2-0.5$~Gyr could have completed a few oscillations if they are bound. The only source with comparable kinematic and cooling age is Gaia DR3 3611573712136684928, but even in this case the cooling age is only one order of magnitude larger than kinematic age. 

\begin{table*}
    \centering
    \begin{tabular}{llcccccccccc}
    \hline
    Name & \textit{Gaia} DR3     & $M_{\rm WD}$  & Cooling age  & Total age & $b$ & $\mu_b$ & $t_\mathrm{kin}$ \\
        & name & (M$_\odot$)   & (Gyr)        & (Gyr)     & ($\circ$) & (mas/year) & (Gyr)\\
    \hline   
    HVWDC 4  & 4739233769591077376  &	 0.54 & 	 0.0223 & 	 5.29 &  -55.997  &  164.491  &    \\
    HVWDC 7  & 855361055035055104 &	 1 & 	     3.17 & 	 3.26 &  51.415  &  221.851  &  0.0008 \\
    HVWDC 8  & 4943575978388814976 &	 0.61 & 	 1.52 & 	 3.15 &  -65.471  &  184.841  &    \\
    HVWDC 11 & 5995439960564759296 &	 0.59 & 	 0.631 & 	 2.81 &  8.647  &  -25.583  &    \\
    HVWDC 24 & 3905910019954089856 &	 0.64 & 	 0.576 & 	 1.84 &  69.348  &  -169.926  &    \\
    HVWDC 10 & 1241636356209099264 &	 0.51 & 	 0.835 & 	 13.4 &  64.394  &  -134.903  &    \\
    HVWDC 13 & 729192473703851264 &	 0.52 & 	 0.264 & 	 10.3 &  59.386  &  -166.109  &    \\
    HVWDC 19 & 3611573712136684928 &	 0.52 & 	 0.203 & 	 9.44 &  50.884  &  12.157  &  0.0151 \\
    HVWDC 22 & 5142197118950177280 &	 0.5 & 	     1.19 & 	 14.3 &  -73.568  &  -787.062  &  0.0003 \\
    HVWDC 23 & 1217609832414369536 &	 0.55 & 	 6.03 & 	 9.83 &  52.274  &  -30.262  &    \\
    \hline
    \end{tabular}
    \caption{The estimated masses, cooling ages and total ages of WD HVS candidates located on the WD cooling sequence, assuming these are Solar metallicity CO-core DA WDs. The WD parameters were estimated using the models of \citet{Bedard_2020}, while the progenitor lifetimes were estimated using the models of \citet{Choi_2016}. $b$ is the Galactic latitude, $\mu_b$ is proper motion in the direction of Galactic latittude and $t_\mathrm{kin}$ is the kinematic age computed as minimal time required to reach the galactic latitude $b$.}
    \label{tab:mass_and_age}
\end{table*}

Eventually we are left with five candidates with estimated total ages significantly shorter than the halo stellar population; HVWDC 4, 7, 8, 11 and 24.
These WDs are therefore prime targets for follow-up spectroscopic observations to constrain their 3D velocity, chemical compositions, and physical properties.

%For objects on the WD cooling sequence we provide estimated ages. We find that most of our candidates have WD cooling ages below 1 Gr, while the rest, besides two, have cooling ages below 2 Gyrs. 

\begin{figure}
    \centering
    \includegraphics[width=\columnwidth]{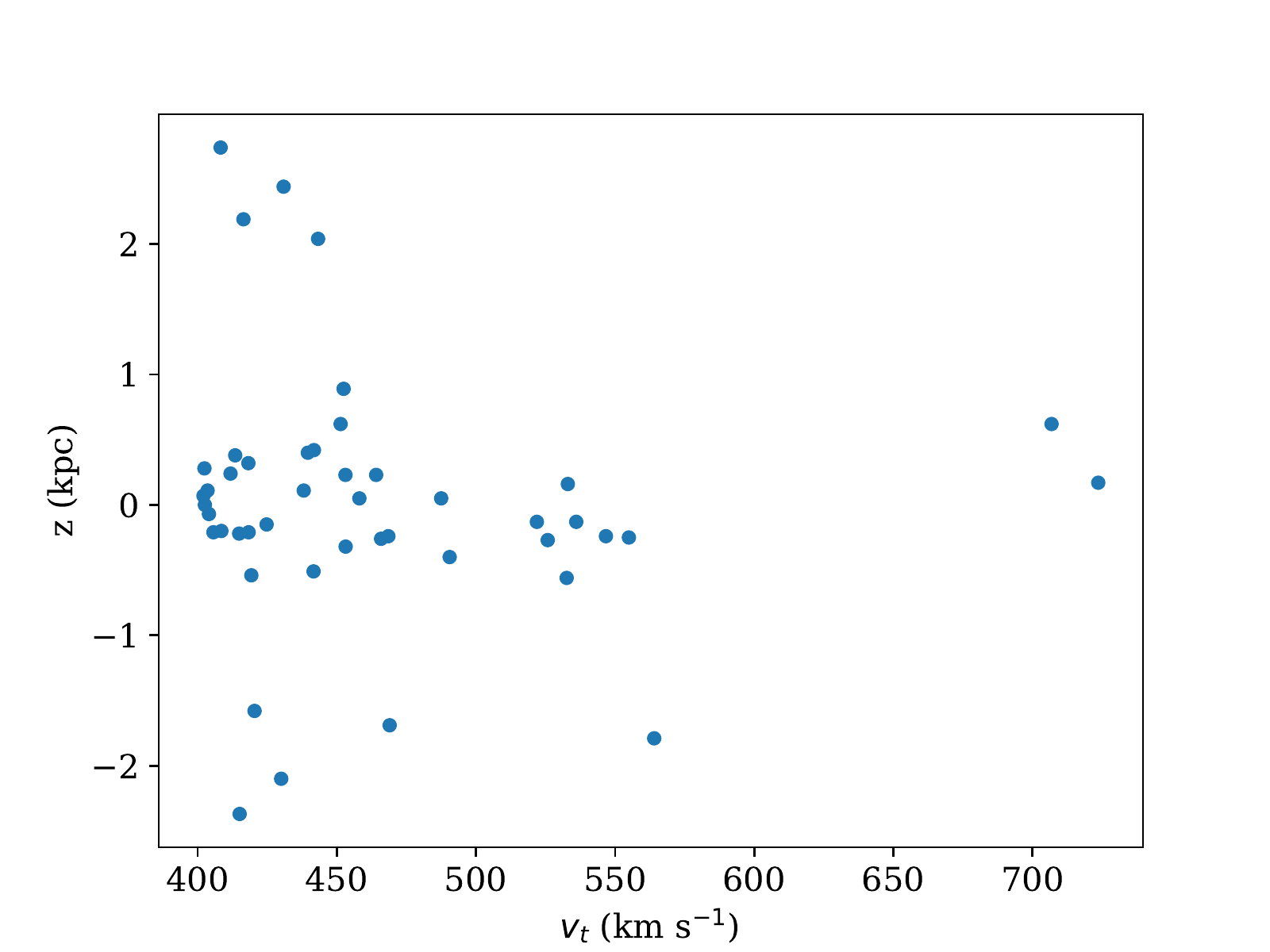}
    \caption{Distribution of transversal velocities and heights above the Galactic plane for hyper-runaway WD candidates. }
    \label{fig:vz}
\end{figure}

In order to potentially provide an alternative age estimates for the other candidates,  we also provide the height of the observed WDs above the disc, see Figure~\ref{fig:vz}. \cite{Cas+16} show that the vast majority of stars residing below 1~kpc from the plane are stars younger than the 10~Gyr age of halo stars. Most of our HVS WD candidates reside below 1~kpc from the plane with the majority below 0.5~kpc (see Table~\ref{tab:candidates_properties}). These findings are consistent with a disc origin for the majority of the sample. One should note that due to their low luminosities, one cannot identify very far WDs, which therefore a priori limits the largest distances, and hence also the heights above the disc.

We conclude that excluding a halo origin for any individual WD (with no age estimate) in our sample is challenging, but that most of our candidate HVS WDs are likely to have a disc origin, and require a non-trivial velocity kick. Nevertheless, radial-velocity follow-ups are required to better constrain/confirm their origin.  

\subsubsection{LSPM J1240+6710/Gaia~DR3 1682129610835350400} 
For this WD we could infer a high mass of $\sim0.79$ M$_\odot$ and a small cooling age of about 0.03~Gyr. In a detailed spectroscopic observations its mass was estimated as 0.41~M$_\odot$ \citep{Gan+20}.
Given that the progenitor mass of such WDs is $3-4$~M$_\odot$ (with MS lifetimes of $\sim400-500$~Myr), such WD cannot be a halo WD, and therefore likely originated in the disc. Its $\sim420$~km\,s$^{-1}$ tangential velocity (and measured 177~km\,s$^{-1}$ radial velocity) therefore makes it a very high velocity runaway, requiring some evolution of dynamical velocity excitation mechanism. Furthermore, it is a oxygen dominated WD. %similar to LP93-21 that was suggested by \cite{Ruffini2019MNRAS} to originate from a type Iax SN. 
Interestingly, this WD has been indeed already identified independently as a potential partly burnt SN remnant by \cite{Gan+20}, due to its peculiar atmospheric composition derived from follow-up observations. Our independent identification of this candidate by its velocity (and exclusion as a halo WD), rather than the previous identification due to its unique spectral features, further supports our approach in identifying potential WDs related to SN explosions.

\begin{table}
\centering
\begin{tabular}{lccccccccc}
\hline
\textit{Gaia} DR3 name & g     & G     & Bp - Rp & R     & z \\
              & (mag) & (mag) & (mag)   & (kpc) & (kpc) \\
\hline
5703888058542880896 & 19.6 & 10.27 & 0.32 & 8.81 & 0.17\\
3688712561723372672 & 18.77 & 9.53 & -0.02 & 8.31 & 0.62\\
6368583523760274176 & 16.87 & 4.31 & -0.37 & 6.78 & -1.79\\
6416314659255288704 & 20.0 & 11.39 & 0.34 & 8.14 & -0.25\\
6841322701358236416 & 19.1 & 11.32 & 0.02 & 8.29 & -0.24\\
5808675437975384320 & 19.8 & 11.37 & -0.03 & 8.14 & -0.13\\
4376935406816933120 & 17.93 & 9.49 & -0.15 & 8.09 & 0.16\\
4739233769591077376 & 19.41 & 10.25 & -0.34 & 8.42 & -0.56\\
6640949596389193856 & 17.88 & 9.09 & 0.29 & 8.02 & -0.27\\
4753345692095875328 & 19.54 & 13.6 & 0.85 & 8.51 & -0.13\\
6777159394645734400 & 19.29 & 10.37 & -0.21 & 8.04 & -0.4\\
855361055035055104 & 17.64 & 13.83 & 0.29 & 8.53 & 0.05\\
6778670265357654656 & 16.16 & 4.08 & -0.42 & 6.54 & -1.69\\
4943575978388814976 & 20.43 & 13.36 & 0.48 & 8.51 & -0.24\\
2463291012727113216 & 19.81 & 12.55 & 0.35 & 8.62 & -0.26\\
1241636356209099264 & 19.66 & 12.6 & 0.32 & 8.4 & 0.23\\
5995439960564759296 & 19.91 & 12.17 & 0.14 & 8.19 & 0.05\\
2737084320170352256 & 18.91 & 10.13 & -0.19 & 8.42 & -0.32\\
2119975000945142272 & 19.98 & 10.93 & 0.03 & 8.4 & 0.23\\
1544331701176666624 & 18.86 & 8.95 & -0.15 & 8.73 & 0.89\\
1470682632777169664 & 19.37 & 10.33 & -0.11 & 8.45 & 0.62\\
3528713077053554432 & 17.2 & 5.04 & -0.42 & 7.72 & 2.04\\
729192473703851264 & 19.82 & 11.39 & -0.07 & 8.73 & 0.42\\
6438915331219654400 & 14.61 & 4.3 & -0.44 & 7.59 & -0.51\\
3537042874067950336 & 17.49 & 8.12 & -0.38 & 8.48 & 0.4\\
1415765359864865408 & 20.21 & 13.77 & 0.79 & 8.47 & 0.11\\
3737057611255721472 & 16.02 & 4.0 & -0.42 & 8.03 & 2.44\\
3195038476578336256 & 16.83 & 4.39 & -0.34 & 10.65 & -2.1\\
4615529846653846016 & 20.09 & 12.79 & 0.63 & 8.39 & -0.15\\
6670029411202563584 & 16.82 & 4.37 & -0.43 & 5.9 & -1.58\\
4925179671389315968 & 19.81 & 10.92 & -0.03 & 8.34 & -0.54\\
2914272062095015552 & 19.9 & 11.38 & 0.08 & 8.8 & -0.21\\
1682129610835350400 & 18.4 & 10.27 & -0.29 & 8.66 & 0.32\\
3948319763985443200 & 17.49 & 5.76 & -0.14 & 8.45 & 2.19\\
2335322500798589184 & 16.52 & 4.63 & -0.48 & 8.25 & -2.37\\
2654214506741818880 & 20.27 & 12.92 & 0.86 & 8.43 & -0.22\\
3611573712136684928 & 19.66 & 11.24 & -0.11 & 8.27 & 0.38\\
4026695083122023552 & 20.36 & 13.44 & 0.71 & 8.55 & 0.24\\
6414789778364569216 & 19.28 & 10.86 & -0.07 & 8.17 & -0.2\\
1378348017099023360 & 16.79 & 3.93 & -0.48 & 7.54 & 2.74\\
2497775064628920832 & 19.17 & 12.02 & 0.2 & 8.67 & -0.21\\
5142197118950177280 & 17.54 & 13.11 & 0.48 & 8.52 & -0.07\\
1217609832414369536 & 20.59 & 14.91 & 1.14 & 8.43 & 0.11\\
5863122429179888000 & 19.58 & 15.19 & 1.32 & 8.46 & 0.0\\
3905910019954089856 & 19.47 & 12.06 & 0.07 & 8.5 & 0.28\\
1212348119518459392 & 19.75 & 15.0 & 1.35 & 8.46 & 0.07\\
\hline
\end{tabular}
\caption{Apparent magnitudes, absolute magnitudes, colours, Galactocentric distances and heights above the Galactic plane for hyper-runaway WD candidates. The Galactocentric distance is computed assuming $R_\odot = 8.5$~kpc.}
\label{tab:candidates_properties}
\end{table}

\subsection{Peculiar hyper-runaway/hypervelocity objects below the main sequence}

\subsubsection{DR6-1, DR6-2, DR6-3}
These objects were the main focus of several dedicated studies  \citep{Shen2018ApJ,Bauer2021ApJ}; here we only briefly remark on these objects. 
An analysis of the three hypervelocity WD donors, assuming they were ejected following a D6-like scenario (given their $>1000$~km\,s$^{-1}$ velocities), suggested that two of the objects (D6-1 and D6-3) had to be massive ($\sim 1\ {\rm M}{_\odot}$), implying that the accretor would have been an even more massive CO WD in a nearly-equal mass ratio binary \citep{Bauer2021ApJ}. The third hypervelocity WD, D6-2, was found to be of a lower mass, $\sim\ 0.4\ {\rm M}_{\odot}$, with looser mass constraints on its accretor. 
As discussed by \cite{Shen2018ApJ}, D6-2 and D6-3 have radial velocities consistent with being $<100$~km\,s$^{-1}$, casting doubt on the interpretation of these stars as hypervelocity stars. Furthermore, only one of the three WDs of \cite{Shen2018ApJ}, D6-2, is considered to have highly significant parallax by \cite{sco18}, serving as a reliable extreme tangential velocity candidate, and even that one can be considered a doubtful high-speed candidate because of its relatively poor astrometric quality parameters \citep{sco18}.

\subsubsection{Additional candidates}
Given our very high velocity cut for the sub-MS candidates, all of the candidates in Table~\ref{tab:candidates2} have velocities comparable or higher than the Galactic escape velocity, making them, or at least most of them, less likely to be tail high velocity halo objects (but, again, with the caveat of  velocity measurement uncertainties), and these are therefore prime candidates for follow-up studies. 

\section{Search for possible white dwarf--supernova remnant association}
\label{s:snr_wd}
Some of the possible origins of HVS WDs are ejections following SN explosions. Such SNe leave SN remnants, that might be observable thousands of years after the explosion. It is therefore possible that some of our hyper-runaway WD candidates might be kinematically related to some supernova remnants (SNR). Such possibility was recently explored by searching for WDs close to SNRs. \citet{Cha+22} studied a SNR proposed to be the counterpart of LP~398-9 (D6-2). \citet{Shi+22} made a deep search for HVS WDs near SNR~1006, with null results. Here we search for all known SNRs whose positions could be consistent with the past propagation of WD HVSs. In order to investigate this possible link we trace back the coordinates of fast-moving WDs from Table~\ref{tab:candidates} and find the distances between these positions and locations of SNRs from the catalogue of \citet{Green2019JApA,Green2009BASI}. 
Mathematically our procedure is as follows. The historical positions of a WD can be written as:
\begin{equation}
\alpha (t) = \alpha - \mu_\alpha t    
\end{equation}
and 
\begin{equation}
\delta (t) = \delta - \mu_\delta t    
\end{equation}
where $\alpha, \delta$ are the present-day right ascension and declination. Here we do not take into account the Galactic gravitational potential because SNR are short-lived structures with ages rarely exceeding $10-20$~kyr. Moreover, all the WDs in our selection are fast-moving objects which are not much affected by Galactic gravitational potential on even Myr timescales.

The current distance between the SNR and the WD can be computed as:
\begin{equation}
D = \sqrt{(\alpha_\mathrm{SNR} - \alpha)^2 + (\delta_\mathrm{SNR} - \delta)^2}    
\end{equation}
Then we introduce two vectors $\vec p$ and $\vec \mu$. The vector $p$ can be written as:
\begin{equation}
\vec p = \left\{\begin{array}{c} (\alpha_\mathrm{SNR} - \alpha) / D \\
(\delta_\mathrm{SNR} - \delta) / D \\
\end{array}\right.    
\end{equation}
and the vector $\mu$ is a unit vector pointing in direction of the proper motion.
The minimal distance between the WD path on the sky and the SNR can be computed as a vector product of $\vec p$ and $\vec \mu$:
\begin{equation}
d = D (p_\alpha \mu_\delta - \mu_\alpha p_\delta).
\end{equation}
We assume that the SNR and the WD are related if $D < S_\mathrm{SNR}$ or $d < S_\mathrm{SNR}$ where $S_\mathrm{SNR}$ is the SNR size as provided by \cite{Green2019JApA}. We additionally introduce the following constrains: (1) The WD should be moving away from the SNR, i.e. $D$ is expect to grow with time, (2) the time necessary for the WD to reach its current position with respect to the SNR should be shorter than 100~kyr. With these constraints only WD Gaia~DR3 5863122429179888000 could originate from a few SNRs. It happens because this WD is located close to the Galactic plane and move along the plane. Among SNR candidates only G309.8+00.0 and G310.8-00.4 seem plausible because they are shell SNR without any clear age estimate. Also, Gaia~DR3 5863122429179888000 has a parallax of 13.23~mas which means that it is located at a distance of $\approx 75$~pc from the Sun, while the distance of G310.8-00.4 is estimated to be 5~kpc. Two other candidates are not plausible because G308.8-00.1 seems to be associated to the radio pulsar PSR J1341-6220, while G315.4-02.3 is young ($\approx 2000$~years) and located at a distance $\approx 2$~kpc.

\begin{table*}
\centering
\begin{tabular}{ccccccc}
\hline
\textit{Gaia} DR3 name  & SNR name & SNR min angular size & $d$ & $t$ \\
               &          &      (arcmin)        & (arcmin) & kyr \\
\hline
5863122429179888000 & G308.8-00.1 & 20.0 & -6.12 & 36658.24\\
5863122429179888000 & G309.8+00.0 & 19.0 & 14.46 & 43040.1\\
5863122429179888000 & G310.8-00.4 & 12.0 & 5.52 & 50656.81\\
5863122429179888000 & G315.4-02.3 & 42.0 & 6.37 & 85085.09\\
\hline
\end{tabular}
\caption{Possible association between hyper-runaway WD and SNRs. $d$ is the minimal angular distance between SNR centre and past trajectory of the star, and $t$ is the time of the closest approach.}
\label{tab:WDSNR}
\end{table*}

\section{Discussion}
\label{sec:discussion}

\begin{table*}
\centering
\begin{tabular}{llclcc}
\hline
Scenario & SNe I type & Speed of ejected WD & Features & Rate \\
               &                 & km\,s$^{-1}$                &                & SNe Ia rate \\
\hline
Double degenerate dynamical detonation (D6) & Ia  & $>1000$        & & 1.0 \\
Hybrid-WD reverse detonation                & Ia  & $1000-1500$    & Heated and slightly polluted WD & 0.01 \\
Failed detonation/weak deflagration model   & Iax & $100$~--~$500$ & Hot, massive, polluted WD & 0.2-0.5  \\   
Single-degenerate double-detonation         & faint Ia & $<600$    & sdB/sdO star is ejected \\
Dynamical ejection in dense collisional enviroments & No SNRe & $<400$ & \\
Binary/triple disruption by massibe black hole in Galactic centre & No SNRe & bound and unbound & Trajectory leads to Galactic centre & 0.3 per year \\
Stripped stars from inspiralling galaxies & No SNRe & bound and unbound & Related to stellar streams \\
Natal kick in a binary with neutron stars & No SNRe & $< 400$ & Tight NS-WD binary & \\
\hline
\end{tabular}
\caption{Summary of different scenarios which could lead to ejection of hyper-runaway WDs (or their immediate progenitors) together with their expected rates and features.\label{t:scenarios}} 
\end{table*}

\subsection{Possible origins of hyper-runaways and hypervelocity WDs}
\label{sec:kicks}
As discussed above, we identify several unbound hypervelocity WDs, which likely require significant velocity kicks, and about twenty HVS WDs and peculiar objects with velocities approaching the Galactic escape velocity. These identifications are very good candidates for WDs which experienced significant velocity kicks (even if formed in the halo). We also identified additional several tens of potential hyper-runaway WDs with lower velocities, which could also arise from such processes but might, instead, belong to the tail of non-kicked halo WDs. 

We summarised all the scenarios leading to hyper-runaway WD formation in Table~\ref{t:scenarios} together with the velocities of produced remnants and their important features. We will briefly go through all of these scenarios and explain if they are compatible with the observations. 

%Velocity excitation of a few hundreds of km\,s$^{-1}$ likely requires an evolutionary or dynamical excitation mechanisms. In the following we briefly review these processes, and discuss our finding in this context. 

\subsection{Velocity kicks from SN explosions}
\subsubsection{Double-degenerate dynamical detonations and ejection of WDs} 

As suggested by \cite{Gui+2010,finketal10}, two CO WDs with helium envelope are driven to each other by emission of gravitational radiation. At some point, after Roche overflow is initiated, the accretor experience thermonuclear detonation which causes secondary detonation inside the CO core of the accreator. The donor WD survives the SN Ia event and receives a large speed comparable to $1000-1500$~km\,s$^{-1}$. 

D6-1--D6-3, the only identified candidates with $>1000$ km\,s$^{-1}$ were suggested to originate from this scenario \citep{Shen2018ApJ,Bauer2021ApJ}. However, the velocity of the best astrometric candidate, D6-2 (and potentially also D6-1 and D6-3, if they lie on the lowest velocity regime of the measurement uncertainty) could also be explained by a very different scenario such as the reverse detonation scenario mentioned above \citep{Pak+21} and further discussed below, rather than the D6 scenario. Finally, even assuming all three hypervelocity WDs are related to the D6 scenario, the non-detection of others in our and previous studies suggests that their ejection rate is at least two orders of magnitudes less than the inferred SN Ia rate \citep{Shen2018ApJ}.
Furthermore, in our whole sample we find only these three candidates to have estimated velocities exceeding $1000$~km\,s$^{-1}$. Even accounting for the 95~per~cent uncertainty intervals we calculated, only seven more candidates could have velocities exceeding $1000$~km\,s$^{-1}$. In other words, even accounting for unlikely (but still possible) very large measurement errors, the overall number of extreme HVSs from a D6-like scenario is very small, compared to the type Ia SN rate expectations. Thus, this scenario is unlike to be responsible for all SN Ia events.

\subsubsection{Hybrid-WD reverse detonation and ejection of WDs}
In alternative scenario suggested by \cite{Pak+21}, the first detonation in the accretor He shell does not trigger the secondary detonation in the core, but instead the burning front propagates back to donor hybrid HeCO WD and its core detonates. Thus, in this scenario the donor is disrupted and accretor receives a large speed. The expected rate is around 1~per~cent of all SNe Ia.
In term of rates, this scenario could therefore potentially explain all the observed $>1000$~km\,s$^{-1}$ hypervelocity WDs, and be consistent with such a small number of identified HVS WDs. As discussed above, the low radial velocity of these objects raises concern regarding the reliability of the extreme velocity measurements. However, taken at face value, the reverse detonation is consistent with D6-2 velocity but at most marginally with the higher velocities of D6-1 and D6-3.   

\subsubsection{Failed-detonation/weak-deflagration model for Iax SNe and ejection of polluted WDs}

As suggested by \cite{Jor+12}, an ignition of nuclear burning might not lead to a full detonation, but it could leave most of the WD intact. This event is seen as faint peculiar SNe Iax and could occur in $20-50$~per~cent of SNe Ia cases.

A number of a few up to a few tens of hyper-runaway WDs (possibly polluted or extremely hot WDs which would likely be observed as peculiar objects) would be comparable to the number of identified HVS WD candidates reported here and their origins could therefore be consistent with type Iax SNe.       

\subsubsection{Single-degenerate double-detonation models for faint Ia SNe and ejection of sdB/sdO stars}

As suggested by \cite{woosleytaamweaver86}, a massive WD accretes material from He-rich stellar companion. This companion becomes an sdB or sdO star. The accretor eventually accumulates enough mass to trigger explosion of the CO core which unbinds the companion.
Our potential finding of tens such candidates, if verified as hyper-runaway sdB/O stars, could then be more consistent with the theoretical estimates of \citet{Neu+22}.

\subsection{Dynamical ejections}

\subsubsection{Binary disruption following core collapse supernova explosion}

Some WDs can be formed as a result of binary disruption where the primary evolves off the main sequence and explodes as a core-collapse supernova. The secondary turns into a runaway star which eventually turns into a runaway WD. The formation of runaway and hyper-runaway stars via binary disruption following core-collapse supernova explosion was studied by \cite{Blaauw1961BAN,Tauris1998AA,PortegiesZwart2000ApJ,Tauris2015MNRAS,Evans2020MNRAS}. In particular, \cite{Tauris2015MNRAS} found that standard binary stellar evolution could lead to ejection of stars with speeds above $v > 400$~km\,s$^{-1}$. However, the number of ejected stars with these velocities is not sufficient to explain hyper-runaway stars seen in our Galaxy \citep{Evans2020MNRAS}. The number of hyper-runaway stars produced via this channel is very sensitive to the common-envelope parameter $\alpha$ and the natal kick velocity distribution, and could be increased if these two parameters are tuned. The rate of hyper-runaway WD formation via this channel was not studied in the literature, but at best it could give $<2$~per~cent of all type-II SNe.

\subsubsection{Dynamical ejections in dense collisional environments}
It was suggested that close encounters between binaries and other stars/binaries in clusters could give rise to energy and momentum exchange leading to the ejection of runaway stars \citep{Leo+90,Perets2012ApJ,Oh2016AA}. Such stars, once they evolve, could later become WDs. However, the ejection velocities are of the order of the orbital velocities of the binary components participating in the encounter, which are typically limited (e.g. \citealt{Leo+90} obtained at most $\sim200$~km\,s$^{-1}$). One of us \cite{Perets2012ApJ} have shown in a detailed study of young clusters that the resulting runaway stars are typically ejected at moderate velocities and that ejection of hyper-runaway stars with velocities exceeding 400~km\,s$^{-1}$ is rare. Ejection of runaway binaries, likely required for the formation of most sdB stars, is at even lower velocities.

\subsubsection{Binary/triple disruptions by the massive black hole in the Galactic centre}
Binary disruption by a massive black hole was suggested to produce hypervelocity stars \citep{Hil88} ejected at hundreds of km\,s$^{-1}$ and even higher velocities, and could be unbound from the Galaxy. Observations of such hypervelocity stars \citep{Bro+05,Ede+05,Bro+15} constrain their total number to at most a few hundreds of B-stars observable in the Galactic halo. Though bound hypervelocity stars could be more abundant \citep{per+09,Gen+22} their numbers would also be limited.
\cite{Evans2022MNRASa,Evans2022MNRASb} found no reliable HVS in \textit{Gaia} DR2 and thus set an upper limit to $3\times 10^{-1}$~years$^{-1}$. 
Nevertheless, bound hypervelocity stars could evolve to become WDs and be accumulated over the lifetime of the Galaxy, contributing to the population of extreme velocity WDs. This can be better constrained by following the WD trajectories and check whether they could be consistent with a Galactic centre origin. This is beyond the scope of this works, but can be explored in later studies. 
While this mechanism can eject binary HVSs, it is less likely to do so, and therefore less likely to eject interacting binaries that might form sdB/O stars \citep{Per09,Per09b}.   

\subsection{Stripped stars from inspiraling galaxies}
\citet{Aba+09,Piffl2011AA} suggested that some hypervelocity stars in the halo could also originate from stars stripped from a current or a past inspiraling galaxy. Such stars might be younger than the halo stellar population, and therefore could masquerade as disc stars ejected at high velocities. This would likely lead to a positional over density and velocity correlations of hypervelocity stars in the sky. Excluding this possibility requires a more detailed study of the distribution of hypervelocity stars across the sky, which is beyond the scope of the current study. 

\subsection{Natal kicks in binary neutron star -- WD systems and the ejection of a NS-WD binary}
Some of the systems summarised in Tables~\ref{tab:candidates} and \ref{tab:candidates2} showing velocities around 500~km\,s$^{-1}$ could potentially be binaries with invisible neutron-star (NS) component. For example, \citet{Heber2009ARAA} discussed that there might exist a hidden population of massive compact companions (NSs or BHs) to sdB stars. 
Some NSs are known to receive natal kicks with amplitudes exceeding $600-800$~km\,s$^{-1}$ \citep{Lyne1994Natur}. If the system was compact before the SN explosion, or the natal kick was orientated favourably, the binary could survive a SN explosion and receive a significant centre-of-mass velocity. 

In a recent work, we modelled different formation channels for NS-WD binaries \citep{Toonen2018AA}. We found that natal kick in the form suggested by \citet{Verbunt2017AA} could produce a small fraction of NS-WD binaries moving with speeds around $350-400$~km\,s$^{-1}$. In this scenario, the NS could be too old to be seen as a radio pulsar. Even if the neutron star is still active as a radio pulsar, its radio beams could miss the Earth for the majority of these objects. NSs are typically very weak optical sources and not expected to be detected by \textit{Gaia}.
However, this hidden NS could be discovered in a series of WD spectral observations because the spectral lines will shift due to the orbital motion of the binary component.  

The maximum systemic velocity which is reached by a NS-WD binary is very sensitive to the natal kick model and the properties of the common-envelope evolution \citep{Toonen2018AA}. Thus, detailed studies of these binaries could help to constrain both these aspects, but generally this channel could lead to the production of hyper-runaway NS-WD binaries, though likely at the lower velocity regime we considered (around 400~km\,$^{-1}$).  A detailed population synthesis is required to estimate the number of HVS WDs with invisible NS which could be seen in the \textit{Gaia} survey.
%\subsection{The origins of HVS WDs}

% Discuss \cite{Neu+22} \cite{Neu+21}
%\cite{Cha+22} D6-1 SNR 
%\cite{Men+21} typical sdB velocities from SNe are low <200 km\,s$^{-1}$ \cite{Gei+15} suggest much higer velocites explaining US 708, but \cite{liu+21} limit this to about 600 km\,s$^{-1}$.

%Farihi et al 2020 suggest mot DQ WDs are not merger products

\section{Conclusions}
In this work we have analysed the \textit{Gaia} DR3 catalogue to search for candidate hyper-runaway and hypervelocity WDs and peculiar objects that could have been ejected at high velocities due to thermonuclear type Ia/Iax SNe or rare dynamical encounters. We identified most of the previously studied candidates (beside a few which had too large measurement uncertainties to pass our quality threshold), and found 46 of {\emph new} WD and other non-MS peculiar HVS candidates (below the MS and above the WD region in the HR diagram). Our new candidates include 4 of highly likely unbound HVS WDs and sub-MS candidates, and additional 42 of possible unbound HVSs (with velocities comparable to the escape velocity from the Galaxy). Among them we identified 25 of hyper-runaway WDs. We  determined the ages of several of these and exclude a halo origin for 5 (HVWDC 4, 7, 8, 11, 24), making them good candidates for being ejected from the Galactic disc through SNe/encounters. Most of the other WD candidates could also originate from the disc, but we cannot exclude a halo origin. 

Overall we find that the number of identified candidates and their velocity distributions could be consistent with the expected contributions from type Iax SNe and reverse detonation of hybrid SNe, but likely rules out the double-detonation D6-model as a main contributor to the origin of normal type Ia SNe. Double-detonation in He-rich single-degenerate models may provide a non-negligible contribution to the origin of sdB/O runaways. 

We also searched for HVS WDs with past trajectories crossing known supernovae remnants, but found only one potential candidate, which might also be a chance coincidence. The lack of more candidates also disfavours the possibility that most type Ia SNe give rise to HVS WDs, as expected in the D6 scenario. 
We encourage follow-up studies of the identified candidates in order to better characterise their velocities and physical properties, which could then provide important constraints on the physical mechanisms for hyper-runaway ejections, and in particular the origins of type Ia/Iax SNe.

\section*{Acknowledgements}
The work of API was supported by STFC grant no.\ ST/W000873/1. HBP acknowledges support for this project from the European Union's Horizon 2020 research and innovation program under grant agreement No 865932-ERC-SNeX. The research of NH is supported by a Benoziyo prize postdoctoral fellowship. We thank anonymous referee, J.J. Hermes, B. G\"{a}nsicke, S.W. Jha and R. Raddi for useful suggestions which helped us to significantly improve the manuscript.

This research has made use of the SIMBAD database,
operated at CDS, Strasbourg, France.

This work has made use of data from the European Space Agency (ESA) mission
{\it Gaia} (\url{https://www.cosmos.esa.int/gaia}), processed by the {\it Gaia}
Data Processing and Analysis Consortium (DPAC,
\url{https://www.cosmos.esa.int/web/gaia/dpac/consortium}). Funding for the DPAC
has been provided by national institutions, in particular the institutions
participating in the {\it Gaia} Multilateral Agreement.

%%%%%%%%%%%%%%%%%%%%%%%%%%%%%%%%%%%%%%%%%%%%%%%%%%
\section*{Data Availability}

This work is based on publicly available data from the \textit{Gaia} archive (see Appendix~\ref{s:adql} for the ADQL query).



\bibliographystyle{mnras}
\bibliography{hvs} % if your bibtex file is called example.bib

%\begin{thebibliography}{99}
%\end{thebibliography}

%%%%%%%%%%%%%%%%%%%%%%%%%%%%%%%%%%%%%%%%%%%%%%%%%%

%%%%%%%%%%%%%%%%% APPENDICES %%%%%%%%%%%%%%%%%%%%%

\appendix

\section{ADQL request}
\label{s:adql}

Our basic request is:
\begin{verbatim}
select top 1000 *, abs(pm) / parallax as v 
from gaiadr3.gaia_source 
where parallax_over_error > 4 and parallax > 0.25 
and RUWE < 1.4 and IPD_FRAC_MULTI_PEAK <= 2 
and IPD_GOF_HARMONIC_AMPLITUDE < 0.1 
and ASTROMETRIC_SIGMA5D_MAX < 1.5 
and PHOT_G_MEAN_MAG - 5*log10(1000.0 / parallax)
+5 > 6 + 5 * bp_rp 
order by v DESC
\end{verbatim}

Our relaxed request is:
\begin{verbatim}
select top 1000 *, abs(pm) / parallax as v 
from gaiadr3.gaia_source 
where parallax_over_error > 4 
and parallax > 0.25 
and RUWE < 1.4 
and IPD_FRAC_MULTI_PEAK <= 2 
and IPD_GOF_HARMONIC_AMPLITUDE < 0.1 
and ASTROMETRIC_SIGMA5D_MAX < 1.5 
and PHOT_G_MEAN_MAG 
- 5*log10(1000.0 / parallax)+5 > 0.66 + 6.67 * bp_rp 
order by v DESC
\end{verbatim}

\section{A posterior estimate for the transversal velocity and credible intervals}
\label{a:posterior}

The nominal transversal velocity is computed as:
\begin{equation}
v_t [\mathrm{km\,s}^{-1}] = \frac{4.74 \sqrt{\mu_\alpha^2 + \mu_\delta^2} [\mathrm{mas\; year}^{-1}]}{ \varpi' [\mathrm{mas}] }     
\end{equation}
If we apply the error propagation technique to this equation we obtain:
\begin{equation}
\sigma_\mu = \sqrt{\frac{\mu_\alpha^2}{\mu_\alpha^2 + \mu_\delta^2} \sigma_{\mu,\alpha}^2 + \frac{\mu_\delta^2}{\mu_\alpha^2 + \mu_\delta^2} \sigma_{\mu,\delta}^2 }  
\end{equation}
\begin{equation}
\sigma_{vt} = v_t \sqrt{\frac{\sigma_\mu^2}{\mu^2} + \frac{\sigma_\varpi^2}{\varpi'^2}}
\label{eq:error_propagation}
\end{equation}
In many cases $\sigma_\mu / \mu < 0.01$ while $\sigma_\varpi / \varpi' \approx 0.2$, thus the parallax uncertainty is the leading contribution to the total velocity uncertainty.
The main problem of this error estimate Eq.~(\ref{eq:error_propagation}) is that it is symmetric around the nominal velocity while symmetric errors of parallax measurement translate to skewed error distribution for distance. This problem is known as the Lutz-Kelker bias \citep{LutzKelker1973PASP} for survey and was addressed in multiple earlier works, see e.g. \cite{BailerJones2015PASP,Igoshev2016AA}.

This problem can be solved if we write a Bayesian posterior for transversal velocity and make all our priors explicit. 
In our approach we assume that proper motion is measured exactly, thus $\mu / \sigma_\mu \gg 1$ which is true for our high-speed objects. For example for \textit{Gaia} DR3 5703888058542880896, $\mu / \sigma_\mu \approx 500$.

Even though our proper motion measurements are so precise, we need to choose a prior for the velocity distribution to specify our expectations about the velocity. It is useful to show that uniform prior is relatively bad assumption. Let us for a moment assume that $v_x \sim U(-v_\mathrm{max}, v_\mathrm{max})$. It means that transversal velocity $v_t = \sqrt{v_x^2 + v_y^2}$ is drawn from another distribution.  In order to obtain a distribution for $v_t$ we transform to a polar coordinate system and integrate over angle $\theta$:
\begin{equation}
p(v_t) = \int_0^{2\pi} U (v_t \sin \theta) U (v_t \cos \theta) v_t d\theta  
\end{equation}
where $U(x)$ is the probability density function for uniform distribution. The result of this integration is counter-intuitive:
\begin{equation}
p(v_t) = \left\{ \begin{array}{cc} v_t / (4 v_\mathrm{max}^2), & \; \mathrm{if}\; |v_x| < v_\mathrm{max} \; \& \; |v_y| < v_\mathrm{max} \\ 
0 & \; \mathrm{otherwise} \end{array} \right.    
\end{equation}
It means that it is more probable for system to have $v_t$ comparable to $v_\mathrm{max}$ than small some velocity. Also increasing $v_\mathrm{max}$ will lead to an increase in speed. Thus such a prior will not be useful. Instead we introduce a prior in form of normal distribution for each velocity component:
\begin{equation}
v_\alpha = \frac{1}{\sqrt{2\pi} \sigma} \exp\left(-\frac{v^2}{2\sigma^2}\right)    
\end{equation}
This prior has multiple great properties: (1) smaller velocities are more probable than larger velocities, (2) if $\sigma$ is large in comparison to velocities under study, the prior becomes similar to uniform prior. The tangential velocity distribution is simply:
\begin{equation}
f_v(v_t) = \int_0^{2\pi} \frac{v_t}{2\pi \sigma^2} \exp\left[-\frac{v^2\sin^2\theta}{2\sigma^2}-\frac{v^2\cos^2\theta}{2\sigma^2}\right] d\theta = \frac{v_t}{\sigma^2} \exp \left[-\frac{v_t^2}{2\sigma^2}\right]    
\end{equation}

Here we replace the component of tangential velocity $v_\alpha$, $v_\delta$ by using absolute value of the velocity $v_t$ and polar angle such a way that:
\begin{equation}
v_\alpha = v_t \sin \theta    
\end{equation}
and 
\begin{equation}
v_\delta = v_t \cos \theta    
\end{equation}
The joint probability to measure parallax $\varpi'$, proper motions $\mu_\alpha'$ and $\mu_\delta'$, tangential velocity $v_t$, polar angle $\theta$ and distance $D$ can be written as:
$$
p(\varpi', \mu_\alpha', \mu_\delta', v_t, \theta, D) = g(\varpi' | D) f_D (D; l, b) g(\mu_\alpha' | v_t, \theta, D) 
$$
\begin{equation}
\hspace{4cm} \times g(\mu_\delta' | v_t, \theta, D) f_v(v_t; \sigma)   
\end{equation}
where $g(\varpi' | D)$ is the conditional probability to measure parallax given the actual distance. It is written as the normal distribution:
\begin{equation}
g(\varpi' | D) = \frac{1}{\sqrt{2\pi} \sigma_\varpi} \exp\left[-\frac{(1/D - \varpi')^2}{2\sigma_\varpi^2}\right]    
\end{equation}
where $\sigma_\varpi$ is the uncertainty of parallax measurement. 
The function $f_D (D; l, b)$ is our Galactic prior for distances same as in \cite{Verbiest2012ApJ}:
\begin{equation}
f_D (D; l, b) = D^2 R^{1.9} \exp \left[ - \frac{|z(D, l, b)|}{h_z} - \frac{R(D, l, b)}{H_R}\right] 
\end{equation}
in this case $l$ and $b$ are Galactic latitude and longitude respectively. Since most star formation occurs in the thin disc we assume that $h=0.33$~kpc and $H_R = 1.7$~kpc using the values by \cite{Lorimer2006MNRAS} found for young radio pulsars.
The functions $g(\mu_\alpha' | v_t, \theta, D)$ and $g(\mu_\delta' | v_t, \theta, D)$ are conditional probabilities to measure respective component of the proper motion given transversal velocity, angle and distance. These are represented by normal distribution in form:
\begin{equation}
g(\mu_\alpha' | v_t, \theta, D) = \frac{1}{\sqrt{2\pi} \sigma_{\mu, \alpha}} \exp\left[- \frac{(\mu_\alpha' - v_t \sin \theta / (4.74 D) - \Delta \mu_\alpha)^2}{2\sigma_{\mu, \alpha}^2} \right]    
\end{equation}
In this equation, $\Delta \mu_\alpha$ is the correction for Galaxy rotation. In this work we do not use these corrections to estimate the credible intervals because our sources are close to the Sun and Galactic rotation does not change their velocities significantly.

In order to get rid of unknown $D$ and $\theta$ we integrate over these quantities. The integral over $\theta$ is hard to compute analytically because it involves terms in form $\propto \exp(-(vt^2 \sin^2 \theta)$. Instead we deal with terms $g(\mu_\alpha' | v_t, \theta, D) g(\mu_\delta' | v_t, \theta, D) $ as the following. The maximum of this function is located approximately at:
\begin{equation}
\tan (\theta + \pi) = \frac{(\mu_\alpha' - \Delta \mu_\alpha)}{(\mu_\delta' - \Delta \mu_\delta)}    
\end{equation}
Sometimes it is shifted by $\pi$ from this location that is why we numerically check both points. 
It is located exactly at this position if $\sigma_{\mu, \alpha} = \sigma_{\mu, \delta}$. 
Thus we use the Nelder-Mead algorithm to iterate and find exact maximum of this function.
Next we iterate again looking for the situation when $g(\mu_\alpha' | v_t, \theta, D) g(\mu_\delta' | v_t, \theta, D)$ decreased by two orders of magnitude. Thus, we end up with two angles $\theta_0$ and $\theta_1$. We integrate numerically between these two angles using 20 mesh points distributed uniformly. The result of this integration is numerical constant $\kappa$ which is unique for $v_t$ and $D$ combination.

\begin{figure}
    \centering
    \includegraphics[width=1.1\columnwidth]{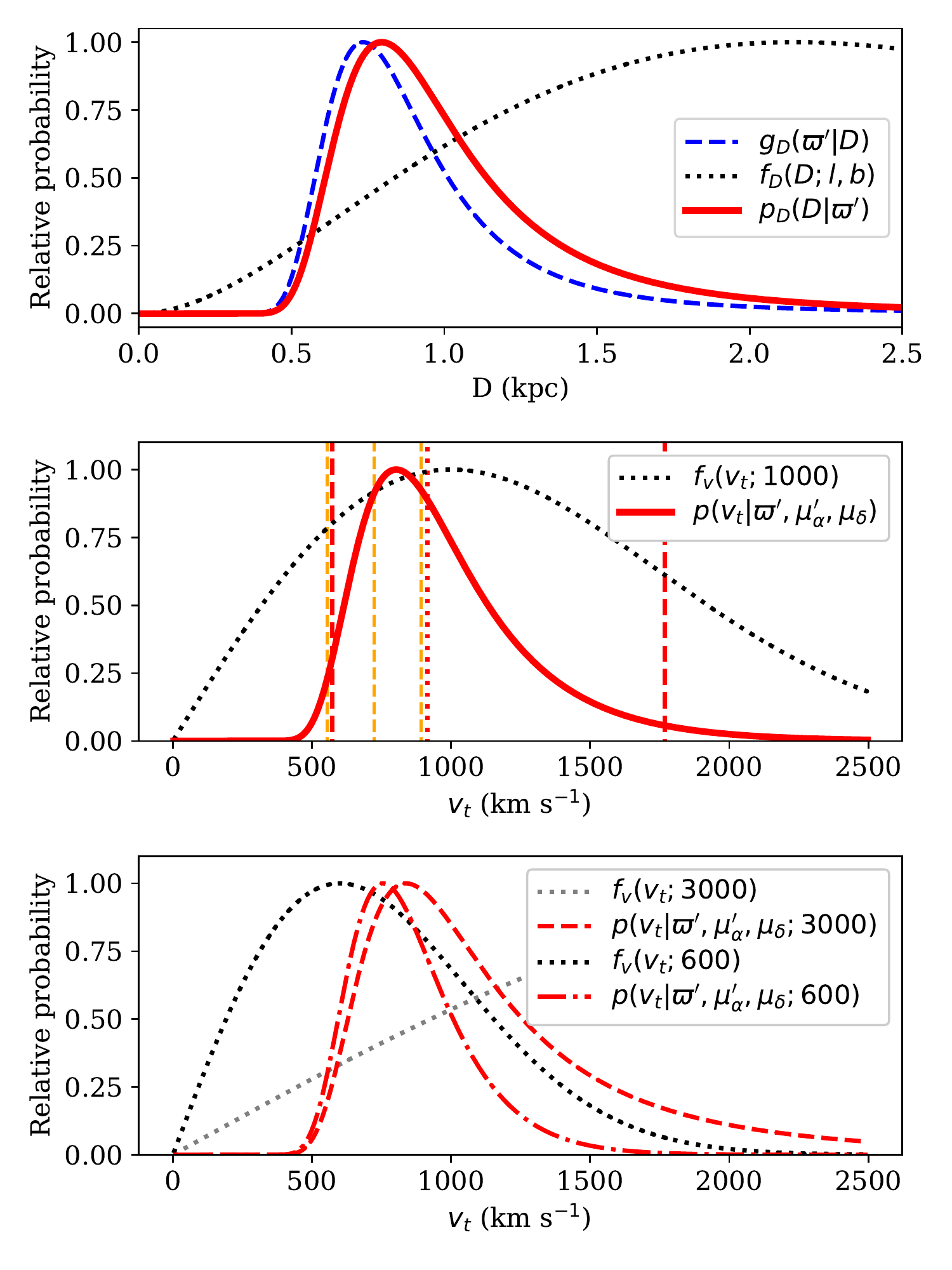}
    \caption{Posterior distribution for distance (top panel) and for tangential velocity (two lower panels) for \textit{Gaia} DR3 5703888058542880896. Dotted black line at the top panel shows the Galactic prior in the direction of the source; dashed blue line shows conditional probability for parallax given distance and solid red curve shows the posterior distribution. Lower panels show the posterior distribution for tangential velocities assuming that proper motion is measured exactly. In the middle panel, yellow lines show the nominal velocity (dashed line) and 68~percent confidence interval computed using the error propagation technique. Red lines show the median of the posterior distribution (dotted line) and 95~percent credential interval. In the lower panel, grey and black dotted lines show priors for velocity distribution with $\sigma = 3000$~km\,s$^{-1}$ and $\sigma = 600$~km\,s$^{-1}$ respectively. Red lines show posterior for the case when it is assumed $\sigma = 3000$~km\,s$^{-1}$ (dashed line) and $\sigma = 600$~km\,s$^{-1}$ (dot-and-dashed line) for the velocity prior.}
    \label{fig:simple_vt}
\end{figure}

Proceeding further we could notice that in our case function $g(\mu_\alpha' | v_t, \theta, D) g(\mu_\delta' | v_t, \theta, D) \approx 1$ only around value:
\begin{equation}
d_1 = \frac{v_t}{4.74 \sqrt{(\mu_\alpha' - \Delta \mu_\alpha)^2 + (\mu_\delta' - \Delta \mu_\delta)^2}} 
\label{eq:d1}
\end{equation}
Basically this is the distance which corresponds to proper motion $\mu$ if $v_t$ is fixed. It happens because our errors for proper motion are tiny. Therefore, we can simply replace the integration over distance with posterior function at distance $d_1$:
\begin{equation}
p(\varpi', \mu_\alpha', \mu_\delta', v_t) = g(\varpi' | d_1) f_D (d_1; l, b)  f_v(v_t; \sigma) \kappa  
\end{equation}
We show the posterior estimate for distance and transversal velocity of Gaia DR3 5703888058542880896 in Figure~\ref{fig:simple_vt}. As it is clear from the figure, error propagation technique gives poor estimate for errors underestimating the size of high-velocity tail. In our calculations we fix $\sigma$ in the velocity prior at value $\sigma = 1000$~km\,s$^{-1}$ which is wide enough to cover all our velocity range. In Figure~\ref{fig:simple_vt}, we also show the posterior distribution for two-velocity computed with prior $f_v(v_t; 3000)$ and $f_v(v_t; 600)$. In the case of $\sigma=3000$~km\,s$^{-1}$ the distribution becomes slightly wider. Respectively, in the case of $\sigma=600$~km\,s$^{-1}$, the posterior velocity distribution shrinks.

%%%%%%%%%%%%%%%%%%%%%%%%%%%%%%%%%%%%%%%%%%%%%%%%%%

% Don't change these lines
\bsp	% typesetting comment
\label{lastpage}
\end{document}